\documentclass[draft]{agujournal2019}
\usepackage{url} 
\usepackage{lineno}
\usepackage[inline]{trackchanges} 
\usepackage{soul}
\draftfalse

\journalname{Geophysical Research Letters}

\begin{document}

\title{FloeNet: A mass-conserving global sea ice emulator that generalizes across climates}

\authors{William Gregory\affil{1}, Mitchell Bushuk\affil{2}, James Duncan\affil{3}, Elynn Wu\affil{3}, Adam Subel\affil{4}, Spencer K. Clark\affil{2,3}, Bill Hurlin\affil{2}, Oliver Watt-Meyer\affil{3}, Alistair Adcroft\affil{1}, Chris Bretherton\affil{3}, Laure Zanna\affil{4}}

\affiliation{1}{Atmospheric and Oceanic Sciences Program, Princeton University, NJ, USA}
\affiliation{2}{Geophysical Fluid Dynamics Laboratory, NOAA, Princeton, NJ, USA}
\affiliation{3}{Allen Institute for Artificial Intelligence (Ai2), Seattle, USA}
\affiliation{4}{Courant Institute School of Mathematics, Computing, and Data Science, New York University, New York, NY, USA}

\correspondingauthor{Will Gregory}{wg4031@princeton.edu}


\begin{keypoints} 
\item FloeNet is a mass-conserving graph neural network sea ice emulator that learns both ice dynamics and thermodynamics
\item FloeNet is trained on a present-day climate and can generalize to pre-industrial and increasing 1\%/year CO\textsubscript{2} climates
\item Emulating budget terms and enforcing mass conservation improves generalization over predicting full state fields
\end{keypoints}

\begin{abstract} 
We introduce FloeNet, a machine-learning emulator trained on the Geophysical Fluid Dynamics Laboratory global sea ice model, SIS2. FloeNet is a mass-conserving model, emulating 6-hour mass and area budget tendencies related to sea ice and snow-on-sea-ice growth, melt, and advection. We train FloeNet using simulated data from a reanalysis-forced ice-ocean simulation and test its ability to generalize to pre-industrial control and 1\% CO\textsubscript{2} climates. FloeNet outperforms a non-conservative model at reproducing sea ice and snow-on-sea-ice mean state, trends, and inter-annual variability, with volume anomaly correlations above 0.96 in the Antarctic and 0.76 in the Arctic, across all forcings. FloeNet also produces the correct thermodynamic vs dynamic response to forcing, enabling physical interpretability of emulator output. Finally, we show that FloeNet outputs high-fidelity coupling-related variables, including ice-surface skin temperature, ice-to-ocean salt flux, and melting energy fluxes. We hypothesize that FloeNet will improve polar climate processes within existing atmosphere and ocean emulators.
\end{abstract}

\section*{Plain Language Summary} 
\noindent Climate models have been used for several decades to make long-term climate projections. These models are very expensive to run, meaning that we typically do not run enough simulations to appropriately constrain uncertainty related to climate change emissions scenarios and internal climate variability. This cost also effectively limits the number of institutions that are contributing to the climate science discourse. In various areas of scientific research, we are seeing how machine learning models can identify patterns in computer model outputs and approximate the underlying physical processes. These machine learning emulators have an advantage that, once trained, they are fast to run, even on a home computer. For climate emulators, this means that users can run hundreds to thousands more experiments than they can with a traditional climate model. This study represents a contribution to the field of climate model emulation. We develop an emulator of one global climate model's sea ice component and show that incorporating our physical understanding of sea ice behavior into the emulator improves its accuracy when evaluated under colder and warmer climates. Our ambition is to combine this sea ice emulator with other atmosphere and ocean emulators to provide a fully machine-learned model of Earth's climate.

\section{Introduction}
Machine learning (ML) models are driving major advances in numerous areas of Earth system science, including remote sensing \cite{Bastin2019,Gregory2024,Chen2025}, climate downscaling \cite{Brenowitz2025,Mardani2025,Perkins2025}, and environmental sensor placement \cite{Andersson2023}. In climate modeling, ML models are now being integrated into large-scale climate model components, leading to improved subgrid parameterization schemes \cite{Mansfield2024,Behrens2025,Zanna2025} and bias-corrected simulations \cite{Watt2021,Chapman2025a,Gregory2026}. However, hybrid and conventional physics-based models remain subject to computational limitations, requiring thousands of CPU cores to achieve modest throughputs of 10-20 simulated years per day. In response to the computational challenges of numerical modeling, we are seeing the rapid development of auto-regressive ML-based climate model emulators. These models have built on the success of data-driven numerical weather prediction models \cite{Pathak2022,Lam2023,Kochkov2024,Allen2025}, now with global-scale climate emulators of the atmosphere \cite{Chapman2025b,Watt2025,Perkins2025}, ocean \cite{Dheeshjith2025}, coupled atmosphere-ocean \cite{Clark2025,Duncan2025}, and Arctic sea ice \cite{Durand2024,Finn2025}. Across these domains, we consistently observe that, once trained, ML emulators can accurately simulate entire Earth system components in a fraction of the time of their numerical model counterparts; this has enabled new research directions, such as the study of extreme events using very large ensembles \cite{Paciorek2025}. 

Sea ice is an integral component of the coupled climate system. By modulating exchanges of heat, moisture, radiation, and momentum between the atmosphere and ocean, sea ice plays a significant role in regulating Earth's surface temperature and high-latitude climate feedbacks \cite{Hahn2021}. For climate emulation, sea ice concentration may be provided as a forcing variable to atmospheric models \cite{Watt2025} or, in a coupled model, emulated prognostically within the ocean \cite{Duncan2025}. Still, some of the largest outstanding surface air temperature biases within these models exist within the polar regions, prompting broader questions about the representation of sea ice within data-driven emulators. Sea ice concentration alone cannot fully capture the complex exchanges of energy, mass, and salt across the atmosphere-ice-ocean interface, and indeed, \citeA{Duncan2025} found better generalization by also prognosing sea ice thickness. We therefore advocate for a sophisticated sea ice emulator which can not only simulate the evolution of sea ice and its response to different forcings, but also provide critical information on coupled exchanges.

To this end, we introduce FloeNet, a data-driven emulator architecture trained on the Geophysical Fluid Dynamics Laboratory (GFDL) global sea ice model, SIS2. FloeNet is a deterministic auto-regressive graph neural network (GNN) which marks a step forward in sea ice emulation as the first model to dynamically evolve sea ice and snow-on-sea-ice while conserving mass and area. Specifically, FloeNet receives mechanical and thermodynamic forcing inputs from the atmosphere and ocean, and predicts ice and snow mass tendencies due to growth, melt, and advection. This yields a mass-conservative and interpretable model, as timestep-to-timestep changes in sea ice area and mass can now be attributed to each term in their respective budget. In this study, we train FloeNet on simulated data from a present-day climate and test its ability to generalize to present-day, pre-industrial control (piControl) and 1\% CO\textsubscript{2} climates. Across these different forcings we show that FloeNet outperforms a non-conservative model at reproducing sea ice trends and inter-annual variability. Our study therefore provides evidence that incorporating physical principles into ML emulators can improve generalization \cite{Falasca2025}.

\section{Data and methods}
\subsection{Training and validation data}
To train FloeNet, we generate a 65-year simulation between 1958--2022 using the GFDL fourth-generation global ocean-sea-ice model, OM4 \cite{Adcroft2019}. OM4 leverages the MOM6 ocean model and SIS2 sea ice model, where both components are configured to a global 0.25$^\circ$ horizontal resolution and forced by JRA-55-do atmospheric reanalysis \cite{Tsujino2018}. We apply post-processing to this simulation by rotating velocity fields to the true east and north directions using the local grid-orientation angle, and then conservatively regridding to a Gaussian 1$^\circ$-resolution latitude-longitude grid; our 1$^\circ$ grid is identical to that of the ACE2 atmospheric emulator \cite{Watt2025} and Samudra ocean emulator \cite{Dheeshjith2025}. We discard the first 11 years of the simulation to allow adequate spin-up of sea ice concentration and thickness. Our final training, validation, and testing periods are 1969--2005 (37 years), 2006--2015 (10 years), and 2016--2022 (7 years), respectively. 

For piControl and 1\% CO\textsubscript{2} validation tests, we rely on atmospheric and oceanic forcing from the GFDL fourth-generation coupled climate model, CM4 \cite{Held2019}. The piControl forcing is from the same simulation used to train the recent coupled SamudrACE emulator \cite{Duncan2025}; we use the last 140 years of this simulation, from 211--350. The 1\% run is initialized from Jan 1 211 of the piControl and run for 140 years with an increasing 1\% per year atmospheric CO\textsubscript{2} concentration. Both piControl and 1\% data follow the same post-processing as the OM4 data described above.

\subsection{FloeNet architecture}\label{sect:arch}
FloeNet is based on the GraphCast model \cite{Lam2023}, with some modifications. We choose this model for its ability to handle grid distortion at the poles by encoding input features onto a multi-resolution icosahedral mesh. Furthermore, GNNs allow us to flexibly set the degree of non-locality within the model. Specifically, we avoid global connections as we do not expect physically-meaningful timestep-to-timestep relationships between, for example, Arctic and Antarctic sea ice. We do however expect some amount of non-locality to be important, given that elastic stresses due to floe interactions propagate rapidly across basin-scales. With these requirements, we restrict the multi-mesh to refinements 4, 5, and 6 of the regular icosahedron (see supplementary section S1) and also only use 4 GNN layers in the processor. We also remove mesh nodes over land.

The atmospheric forcing data used to drive FloeNet are 6-hour averages of downward shortwave and longwave radiation, sensible and latent heat fluxes, zonal and meridional wind stress over sea ice, and the snowfall rate. The ocean forcing data are 5-day averages of ocean-to-ice heat flux, sea-surface height, and zonal and meridional ice-ocean stresses. Prognostic variables are 6-hour snapshots of ice concentration, ice mass, snow mass, and ice-surface skin temperature, as well as 6-hour averages of growth, melt, and transport mass and area fluxes for ice and snow, and ice-to-ocean salt flux. We also output top and bottom melting energy fluxes as diagnostic variables (see Table S1 for a full list of variable names). We choose 6-hour atmospheric forcing and 5-day ocean forcing because these are the current timesteps of the atmosphere and ocean components in the SamudrACE coupled emulator. This means that FloeNet can be readily integrated into SamudrACE in future work. 

Given a 6-hour sea ice timestep, our training period provides 54,056 samples to optimize FloeNet's $\sim$3 million weights. Training for 50 epochs takes $\sim$5 days on 4 H200 NVIDIA GPUs. A 140-year rollout then takes $\sim$4.75 hours on 1 L40S NVIDIA GPU. We refer the reader to supplementary section S1 for further details of FloeNet's architecture and training procedure, and section S2 for results of FloeNet's sensitivity to random seed, mesh resolution, input forcing, time stepping, and architecture.

In section \ref{sect:results} we compare FloeNet to a baseline architecture, which we hereafter refer to as the ``full-state'' model. The full-state model uses the same forcing variables and GNN-based architecture as FloeNet, but rather than emulating mass and area budget tendencies, the prognostic variables are ice concentration, ice mass, snow mass, zonal and meridional ice velocities, and ice-to-ocean salt flux. We choose this as our baseline to understand the impact of our mass-conservation budget-based approach on out-of-sample generalization. The full-state model also has a 6-hour timestep and is trained for 50 epochs.

\begin{figure}[t!]
    \centering
    \includegraphics[width=\linewidth]{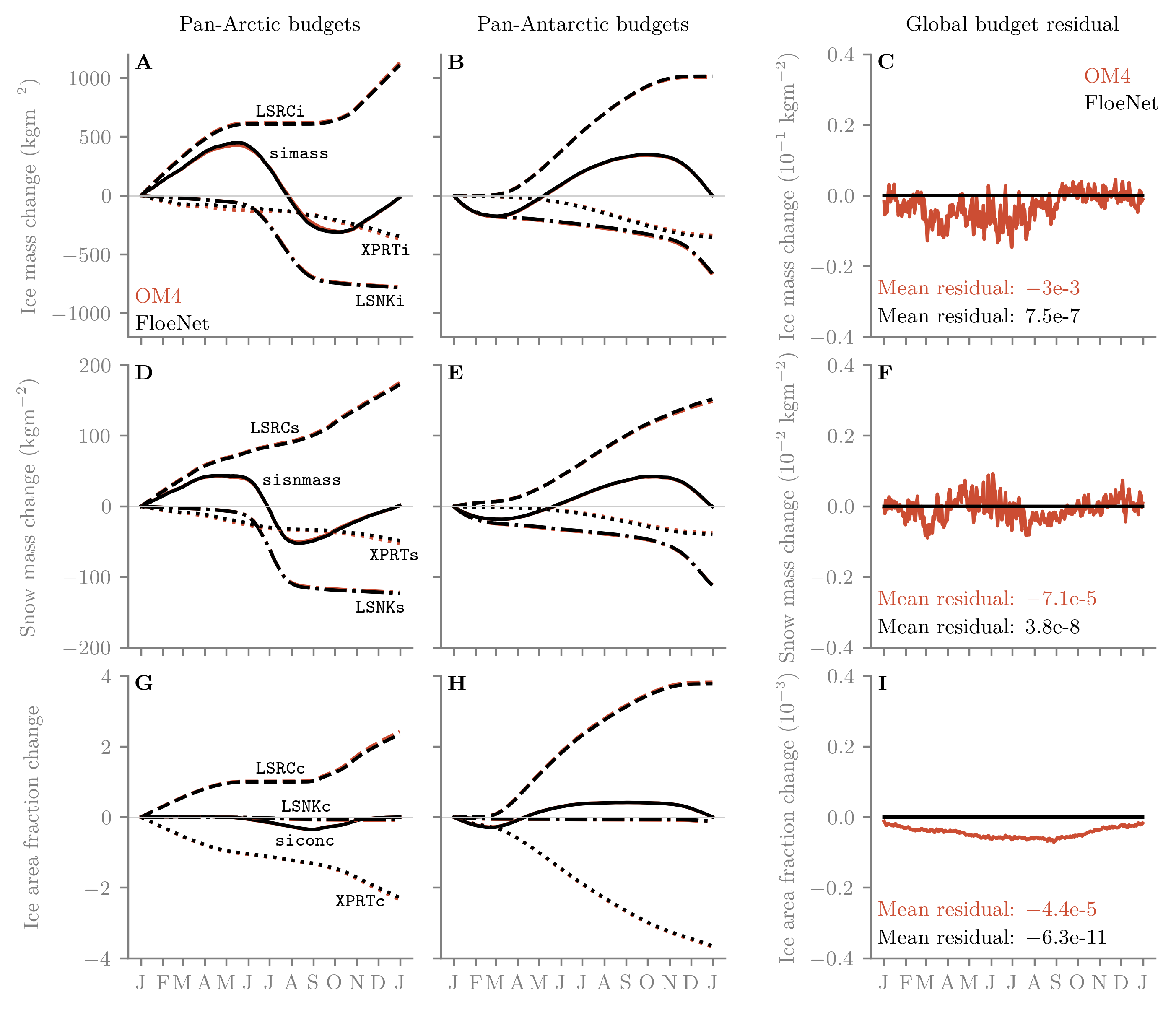}
    \caption{FloeNet mass and area budget anomaly climatologies (relative to Jan 1 00:00:00) for the period 2006--2022. (\textbf{A}) Arctic sea ice mass tendencies due to source terms (\texttt{LSRCi}), sink terms (\texttt{LSNKi}), and transport convergence (\texttt{XPRTi}). The sum of these terms gives the total sea ice mass (\texttt{simass}). (\textbf{B}) Same as (\textbf{A}) but for Antarctic. (\textbf{C}) The global sea ice mass budget conservation error, computed as the difference between the prognostic ice mass and the mass computed through a budget reconstruction. (\textbf{D}--\textbf{F}) Same as (\textbf{A}--\textbf{C}) but for snow mass. (\textbf{G}--\textbf{I}) Same as (\textbf{A}--\textbf{C}) but for sea ice concentration.}
    \label{fig:budgets}
\end{figure}

\section{Results}\label{sect:results}
\subsection{Closing the mass and area budget}
In SIS2, the sea ice mass (\texttt{simass}) evolution equation can be summarized as the change in gridcell-mean ice thickness, $h_i$, due to thermodynamic processes driving ice melt and growth at the top and bottom of the ice, as well as thickness changes due to ice dynamics:
\begin{equation}
\frac{\partial \texttt{simass}}{\partial t} = \rho_i\left.\frac{\partial h_i}{\partial t}\right|_{top} + \rho_i\left.\frac{\partial h_i}{\partial t}\right|_{bot} - \rho_i\nabla\cdot(h_i\mathbf{u}),
\label{eq:budget1}
\end{equation}
where $\mathbf{u}$ is the ice velocity and $\rho_i$ is the ice density (a constant in SIS2). We can further isolate the positive and negative contributions to the thermodynamic terms in (\ref{eq:budget1}), yielding local sources and sinks of ice mass, respectively:
\begin{equation}
\frac{\partial \texttt{simass}}{\partial t} = \texttt{LSRCi} + \texttt{LSNKi} + \texttt{XPRTi}.
\label{eq:budget2}
\end{equation}
Here, \texttt{LSRCi} is the source contribution from ice growth (including congelation growth, frazil growth, and snow-to-ice conversion), \texttt{LSNKi} is the sink contribution from melt (including top melt, bottom melt, evaporation and sublimation), and \texttt{XPRTi} is the ice mass transport convergence. Subsequently, the \texttt{simass} at time $t$ is obtained by temporally integrating (\ref{eq:budget2}) as follows:
\begin{equation}
\texttt{simass}(t) = \texttt{simass}(0) + \int_{0}^t (\texttt{LSRCi}(t) + \texttt{LSNKi}(t) + \texttt{XPRTi}(t)) \ dt.
\label{eq:budget3}
\end{equation}
FloeNet emulates \texttt{simass}, \texttt{LSRCi}, \texttt{LSNKi}, and \texttt{XPRTi} prognostically. To ensure that (\ref{eq:budget3}) is always satisfied, we therefore overwrite FloeNet's prediction of \texttt{simass}$(t)$ with the sum of its predicted budget terms at every timestep, while also ensuring that \texttt{simass} is non-negative. We follow the same procedure for snow mass (\texttt{sisnmass}) and ice concentration (\texttt{siconc}) using their respective budget terms; see supplementary section S1 for more details.

\begin{figure}[t!]
    \centering
    \includegraphics[width=0.8\linewidth]{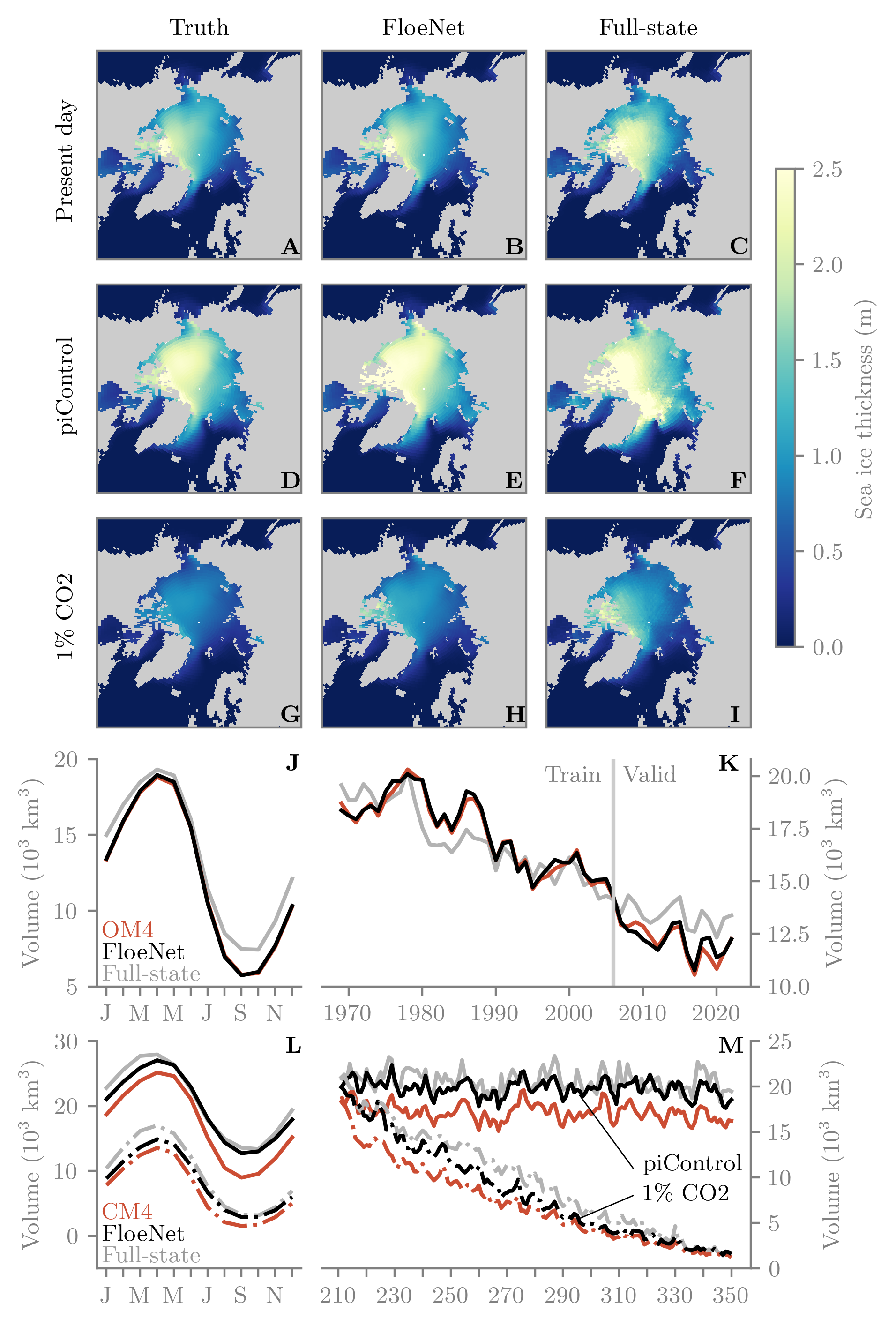}
    \caption{Arctic sea ice thickness and volume mean state and trends. (\textbf{A}) Time-mean sea ice thickness climatology from OM4, computed over 2006--2022. (\textbf{B}) Same as (\textbf{A}) but for a FloeNet rollout with OM4 forcing. (\textbf{C}) Same as (\textbf{A}) but for a full-state model rollout with OM4 forcing. (\textbf{D}) Same as (\textbf{A}) but for CM4 under piControl forcing. (\textbf{E}--\textbf{F}) Same as (\textbf{B}--\textbf{C}) but with CM4 piControl forcing. (\textbf{G}--\textbf{I}) Same as (\textbf{D}--\textbf{F}) but with 1\% CO\textsubscript{2} forcing. (\textbf{J}) Seasonal climatology of Arctic sea ice volume, computed over 2006-2022. (\textbf{K}) Annual-mean Arctic sea ice volume between 1969--2022. (\textbf{L}) Same as (\textbf{J}) but for piControl and 1\% forcing runs, computed over 211--350.  (\textbf{M}) Same as  (\textbf{K}) but for piControl and 1\% forcing between 211--350.}
    \label{fig:ForcingTest}
\end{figure}

Fig. \ref{fig:budgets} shows the FloeNet (emulator) and OM4 (truth) climatologies of \texttt{simass}, \texttt{sisnmass}, and \texttt{siconc} change, and their respective budget contributions. The \texttt{simass}, \texttt{sisnmass}, and \texttt{siconc} curves are computed as follows: we first do a continuous rollout of FloeNet over the validation and test period of Jan 1 2006 00:00 to Dec 31 2022 18:00, using forcing data from the OM4 simulation. Then, for each year between 2006--2022, we subtract the mass (or area) on Jan 1 00:00 from all subsequent timesteps in that year to produce a mass (or area) anomaly time series. We then compute a climatology based on all years. For individual budget terms in Fig. \ref{fig:budgets}, we integrate (\ref{eq:budget3}) from zero mass (or area) using only that term, to show its respective contribution to the mass and area evolution over the course of the year. For Arctic metrics, we average quantities over an area comprising the central Arctic Ocean and Beaufort, Chukchi, East Siberian, Laptev, Kara, and Barents seas (following the domain used by \citeA{Keen2021}). For the Antarctic, we average over grid cells south of 63$^\circ$S. In Fig. \ref{fig:budgets}, we notice that the curves for FloeNet (black) and OM4 (orange/red) are largely overlapping, showing that FloeNet accurately captures the seasonal evolution of sea ice and snow growth, melt, and advection. The budget residuals (Figs. \ref{fig:budgets}C,\ref{fig:budgets}F,\ref{fig:budgets}I) compare the difference between the prognostic \texttt{simass}, \texttt{sisnmass}, and \texttt{siconc}, and those computed via their budgets using (\ref{eq:budget3}), for both FloeNet and OM4. This shows that the FloeNet budget closes to within single-precision roundoff, while OM4 has some small residuals, particularly in \texttt{simass}. In conclusion, Fig. \ref{fig:budgets} highlights the interpretable nature of FloeNet and its ability to learn the seasonal evolution of \texttt{simass}, \texttt{sisnmass}, \texttt{siconc}, and their budget contributions over this 2006--2022 validation and test period.

\subsection{Generalization of mean state, trends, and variability}\label{sect:forcing}
For data-driven climate emulators to be considered trustworthy, they must extrapolate beyond their training distribution. Sea ice is often seen as a barometer for climate change, as the observational record has shown a precipitous downward trend in Arctic sea ice over the last 5 decades \cite{Notz2016}. In this section, we therefore investigate how well FloeNet reproduces sea ice trends, variability, and mean state under present-day, piControl, and 1\% CO\textsubscript{2} forcing. We focus on sea ice thickness and volume, as these are inherently more difficult quantities for an emulator to learn than sea ice concentration and extent, which can be inferred relatively directly from the atmosphere and ocean surface temperatures. We include Figs. S1 and S2 to show that FloeNet reproduces Arctic and Antarctic sea ice extent mean state and trends with high accuracy under different forcings.

Fig. \ref{fig:ForcingTest} shows spatial climatology maps of Arctic sea ice thickness, as well as seasonal climatologies and annual-mean time series of sea ice volume. The present-day truth (Fig. \ref{fig:ForcingTest}A) corresponds to our OM4 simulation over the validation and test period, 2006--2022. Meanwhile, the piControl truth (Fig. \ref{fig:ForcingTest}D) and 1\% CO\textsubscript{2} truth (Fig. \ref{fig:ForcingTest}G) correspond to the CM4 simulations over simulation years 211--350. Comparing FloeNet and the full-state model under present-day forcing (Fig. \ref{fig:ForcingTest}B vs \ref{fig:ForcingTest}C), we can see that FloeNet's sea ice thickness spatial pattern more closely resembles OM4 than the full-state model. The full-state model produces ice which is too thick over the central Arctic and too thin in the Canadian Arctic Archipelago (CAA). We also notice that the full-state model contains a non-negligible amount of grid-scale noise, which may be due to imprinting of the fast-timescale velocity information on the slower-evolving sea ice thickness fields, as seen in ocean emulators \cite{Dheeshjith2025}. The full-state model's over-estimation of sea ice thickness is also reflected in the sea ice volume seasonal climatology (Fig. \ref{fig:ForcingTest}J), which shows a year-round mean volume bias of $+1175$ km$^3$. Meanwhile, FloeNet shows very little bias ($+40$ km$^3$). In terms of present-day trends (Fig. \ref{fig:ForcingTest}K), the full-state model slightly under-estimates the magnitude of the annual-mean sea ice volume trend between 1969--2022 ($-132$ km$^3$/year compared to OM4's $-167$ km$^3$/year). In contrast, FloeNet accurately reproduces the trend across this period ($-166$ km$^3$/year).

Under piControl forcing, FloeNet shows a consistent sea ice thickness spatial pattern with CM4 (Fig. \ref{fig:ForcingTest}D vs \ref{fig:ForcingTest}E), although it has a positive mean-state bias, particularly in the East Siberian, Chukchi and Beaufort seas, and the CAA (see Fig. S3 for bias maps). This bias is also reflected in the sea ice volume seasonal climatology (Fig. \ref{fig:ForcingTest}L; solid curves), with an average year-round value of $+2645$ km$^3$. The full-state model (Fig. \ref{fig:ForcingTest}F) shows a different spatial pattern to CM4, with thick ice north of Greenland extending into the Fram Strait and north of Svalbard, and thin ice towards the Pacific sector of the Arctic. This compensating spatial bias pattern (see Fig. S3E) results in the full-state model having only a slightly larger pan-Arctic sea ice volume bias than FloeNet, at $+3470$ km$^3$.

Under 1\% CO\textsubscript{2} forcing, both FloeNet and the full-state model show a relatively homogeneous sea ice thickness spatial pattern, similar to CM4 (Figs. \ref{fig:ForcingTest}G, \ref{fig:ForcingTest}H, \ref{fig:ForcingTest}I). However, FloeNet has a slight positive sea ice thickness bias, particularly in the CAA. The full-state model then has a larger thickness bias in the area north of Greenland and the CAA. Overall, FloeNet and the full-state model have mean year-round volume biases of $+1389$ km$^3$ and $+2616$ km$^3$, respectively. Interestingly, FloeNet and the full-state model show similar downward trends in Arctic sea ice volume between 211--350, although FloeNet shows visibly lower errors between years $\sim$235--330. CM4 has a trend value of $-1039$ km$^3$/year, while FloeNet and the full-state model have trends of $-1258$ km$^3$/year and $-1335$ km$^3$/year, respectively. The majority of this trend bias occurs within the first 20 years of the run, where CM4 shows a steep initial decline which then levels out and progresses somewhat linearly. We note that both FloeNet and the full-state model stably reach ice-free summers ($< 1$ million km$^2$ of September sea ice extent) under 1\% CO\textsubscript{2} forcing (see Fig. S4). We also highlight that FloeNet has very accurate Antarctic sea ice volume and global snow-on-sea-ice volume under all three forcing tests, while the full-state model has large snow biases in piControl rollouts (see Figs. S4--S7).

\begin{figure}[t!]
    \centering
    \includegraphics[width=\linewidth]{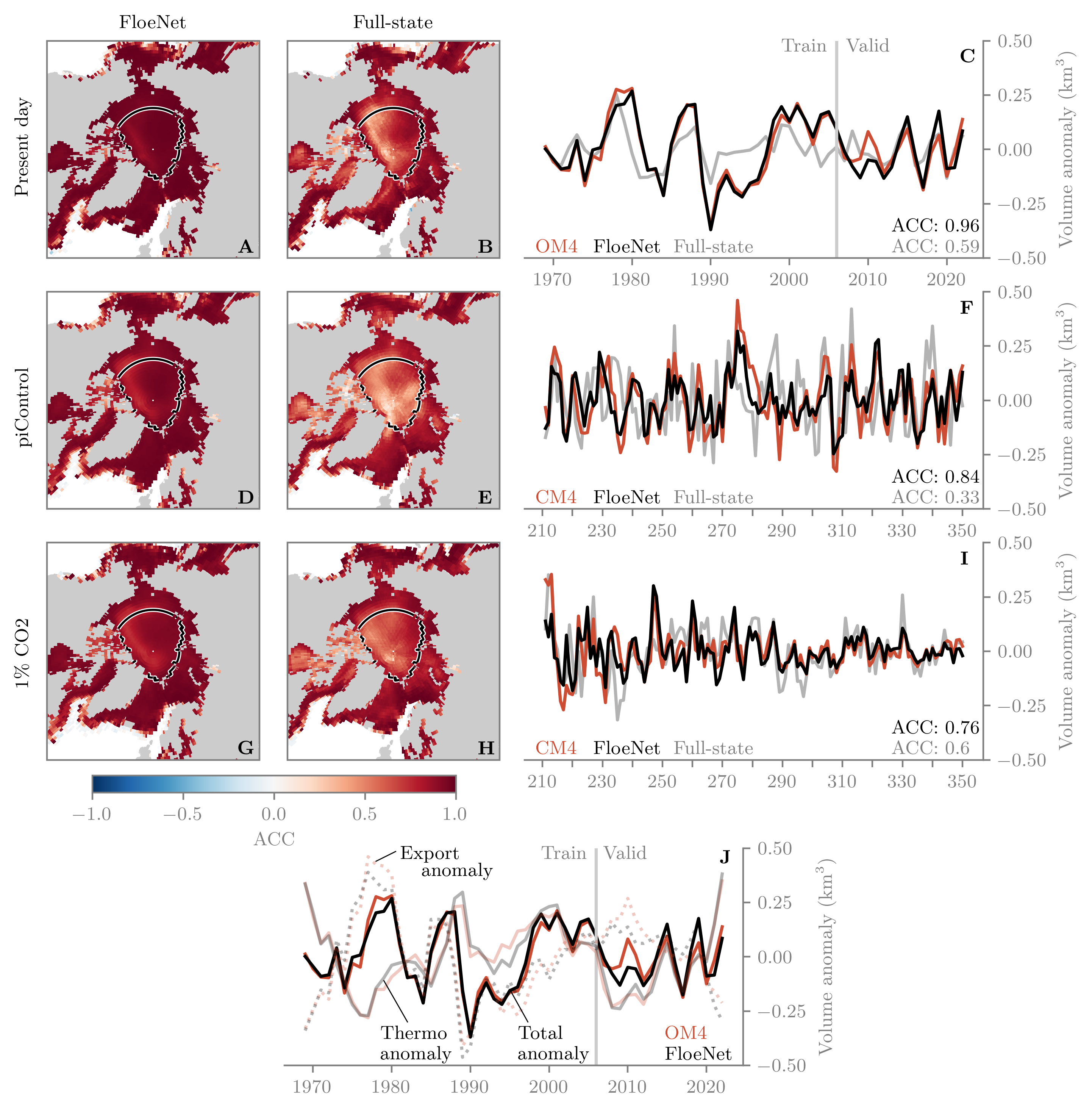}
    \caption{Arctic sea ice volume anomalies under different forcings. (\textbf{A}) Anomaly correlation coefficient (ACC) of annual-mean sea ice volume anomalies between OM4 and FloeNet, computed at each grid cell over the period 1969--2022. (\textbf{B}) Same as (\textbf{A}) but ACC computed between OM4 and the full-state model. (\textbf{C}) Central Arctic annual-mean sea ice volume anomalies over the period 1969--2022. (\textbf{D}--\textbf{F}) Same as (\textbf{A}--\textbf{C}) but for piControl forcing. (\textbf{G}--\textbf{I}) Same as (\textbf{A}--\textbf{C}) but for 1\% CO\textsubscript{2} forcing. (\textbf{J}) Same as (\textbf{C}) but now decomposing FloeNet and OM4 volume anomalies into thermodynamic and dynamic contributions. All central Arctic anomaly time series are computed as averages over the region shown by the contours in the spatial plots.}
    \label{fig:Anoms}
\end{figure}

\begin{figure}[t!]
    \centering
    \includegraphics[width=\linewidth]{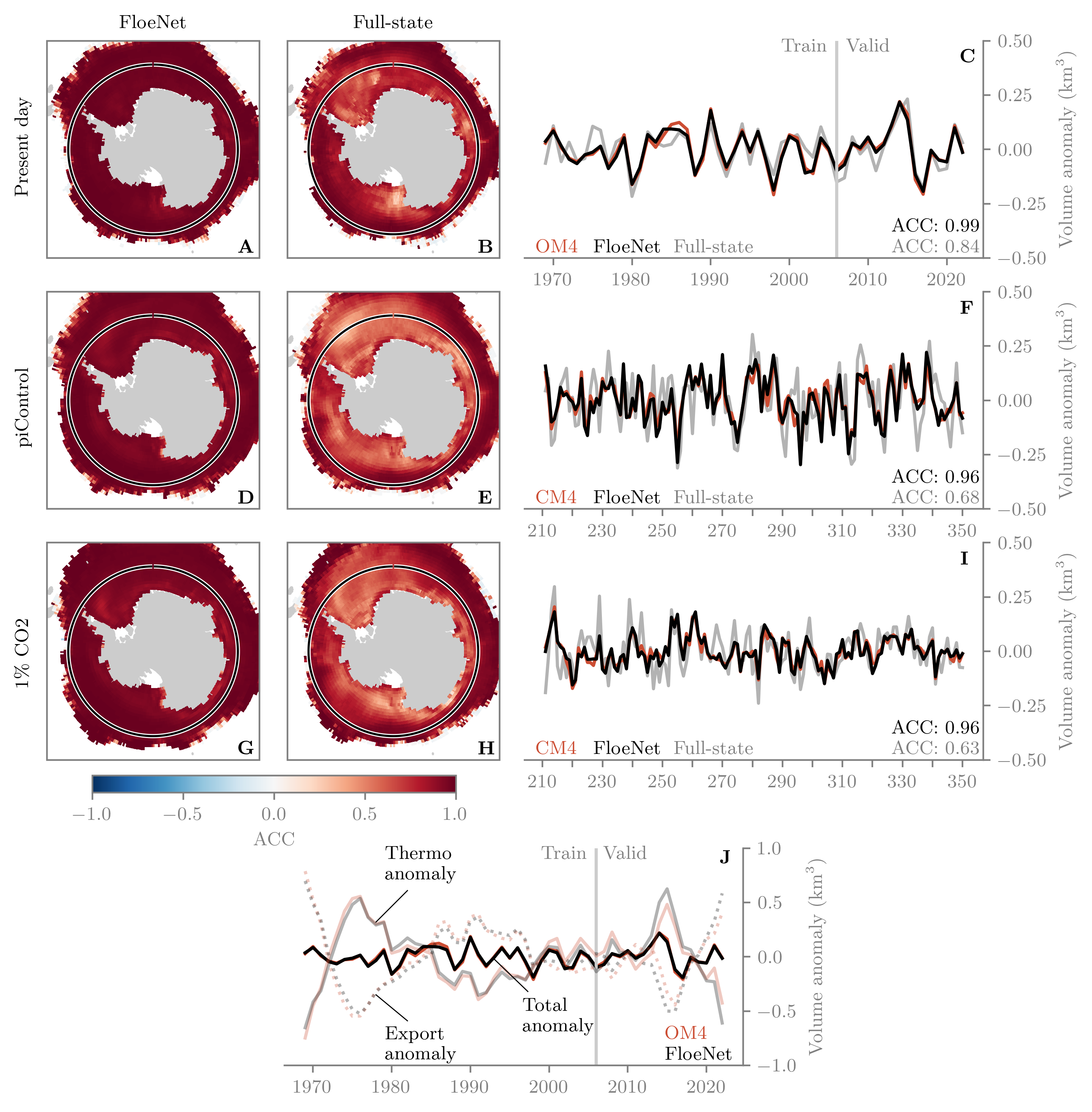}
    \caption{Same as Fig. 3, but for Antarctic. The contour in the spatial maps corresponds to the latitude 63$^\circ$S.}
    \label{fig:AnomsAA}
\end{figure}

We now look at sea ice volume variability. Fig. \ref{fig:Anoms} shows detrended annual-mean Arctic sea ice volume correlations for each model, relative to OM4 (Figs. \ref{fig:Anoms}A and \ref{fig:Anoms}B), CM4 piControl (Figs. \ref{fig:Anoms}D and \ref{fig:Anoms}E), and CM4 1\% CO\textsubscript{2} (Figs. \ref{fig:Anoms}G and \ref{fig:Anoms}H), respectively. FloeNet shows very high pan-Arctic correlations for all three forcing tests. The full-state model also shows high correlations, although they are, on average, lower than FloeNet, particularly within the central Arctic Ocean (contour in Fig. \ref{fig:Anoms}). In Figs. \ref{fig:Anoms}C, \ref{fig:Anoms}F, and \ref{fig:Anoms}I, we compute the mean sea ice volume anomaly within this central Arctic region, where it is clear that FloeNet is the superior model at capturing inter-annual variability in sea ice volume (see anomaly correlation coefficient (ACC) values within each figure panel). For the present-day forcing test (Fig. \ref{fig:Anoms}C), the full-state model shows too little variability, while in the piControl (Fig. \ref{fig:Anoms}F) and 1\% CO\textsubscript{2} forcing tests (Fig. \ref{fig:Anoms}I), it shows too much variability. To better understand FloeNet's high skill, we can use its interpretable budget-based formulation to decompose the Central Arctic volume anomalies into export-driven (\texttt{XPRTi}) and thermodynamic-driven (\texttt{LSRCi}+\texttt{LSNKi}) components. We show this for the present-day (OM4) forcing test, as the CM4 piControl and 1\% CO\textsubscript{2} runs were generated prior to this study and the relevant budget diagnostics were not saved. In Fig. \ref{fig:Anoms}J we can see that FloeNet's export (dashed lines) and thermodynamic (semi-transparent lines) anomalies track the respective OM4 anomalies well, and we can discern from the total anomaly field whether anomalies for a given year were driven by dynamics, thermodynamics, or a combination of both. This gives confidence that FloeNet has correctly learned how to attribute dynamic and thermodynamic processes to changes in sea ice volume.

Fig. \ref{fig:AnomsAA} shows volume anomaly correlations for the Antarctic. In this case, we compute the mean anomaly time series over the region south of 63$^\circ$S. The FloeNet correlations for the piControl (Fig. \ref{fig:AnomsAA}F) and 1\% CO\textsubscript{2} run (Fig. \ref{fig:AnomsAA}I) are very high, with ACC of 0.96 in both cases. This aligns with our previous analysis of sea ice thickness and volume mean state and trends, where we found that FloeNet generally performs better in the Antarctic than the Arctic. In any case, Figs. \ref{fig:ForcingTest}--\ref{fig:AnomsAA} highlight that, while FloeNet may exhibit modest mean-state biases under different forcings, it can accurately reproduce the sign and magnitude of sea ice volume anomalies in both hemispheres. In section \ref{sect:disc}, we outline potential ways to improve the mean-state biases in piControl and 1\% CO\textsubscript{2} rollouts.


\subsection{Potential to improve atmosphere and ocean emulators}\label{sect:fluxes}
FloeNet has been designed to improve atmosphere and ocean emulators by providing output variables on atmosphere-ice-ocean fluxes. We have already shown in the previous two sections that FloeNet can produce high-fidelity mass flux information related to ice growth and melt, which can, in principle, be passed to an ocean emulator to better constrain the water mass budget. In this section, we provide a summary of FloeNet's coupling-related variables: ice-surface skin temperature (\texttt{TS}), ice-to-ocean salt flux (\texttt{SALTF}), top melting energy flux (\texttt{TMELT}), and bottom melting energy flux (\texttt{BMELT}); see supplementary section S1 for exact definitions of \texttt{TMELT} and \texttt{BMELT} and a discussion on surface albedo.

FloeNet and the full-state model produce very accurate seasonal cycles of all variables (see Figs. S8 and S9). However, the full-state model contains a noticeable summertime \texttt{BMELT} bias in both hemispheres. Assuming that the models have learned some relationship between ice thickness and the energy needed for melting, this \texttt{BMELT} bias may be linked to the full-state model's mean-state sea ice thickness bias, which is largest in summer. In Figs. S8 and S9 we also show example snapshots, where both FloeNet and the full-state model are in very good agreement with OM4. Over the 2006--2022 period, the global RMSEs of FloeNet for \texttt{TS}, \texttt{SALTF}, \texttt{BMELT} and \texttt{TMELT} are 0.85$^\circ$C, 6.9$\times 10^{-8}$ kgm$^{-2}$s$^{-1}$, 59.1 Wm$^{-2}$ and 3.7 Wm$^{-2}$, respectively. For the full-state model, RMSEs are 0.83$^\circ$C, 7.6$\times 10^{-8}$ kgm$^{-2}$s$^{-1}$, 64.1 Wm$^{-2}$ and 3.9 Wm$^{-2}$. In supplementary section S2 and Fig. S17, we show that much of the skill in reproducing, for example, \texttt{TS}, comes from our 6-hour timestep, skill which we do not achieve with a coarser 5-day timestep. While this may be due to the significant increase in training samples with 6-hourly data, we would also expect improved generalization from exposing the network to tighter coupling of the sea ice to short-term atmospheric dynamics.

\section{Discussion and conclusions}\label{sect:disc}

In this study, we introduce a mass-conserving global sea ice emulator, FloeNet, that emulates the SIS2 sea ice model. We have shown that, even though trained on simulated data from a reanalysis-forced ice-ocean simulation, FloeNet produces accurate sea ice and snow-on-sea-ice trends and variability under a pre-industrial control (piControl) climate and a climate with increasing 1\% per year atmospheric CO\textsubscript{2} concentration.  Compared to a baseline non-conserving model, both models produce accurate sea ice extent trends and variability and similar pan-Arctic sea ice volume mean states under piControl and 1\% CO\textsubscript{2} climates. However, the ice thickness spatial bias patterns (Fig. S3) reveal that the non-conserving model contains grid-scale noise and large compensating biases, which are otherwise not seen within integrated metrics. In terms of sea ice volume variability, FloeNet is by far the superior model. Additionally, the mass-conservative formulation enables a physical interpretation of emulator errors, analogous to techniques for error attribution in conventional numerical models. Through this framework, we show that FloeNet produces the correct physical response to forcing, accurately partitioning thermodynamically- and dynamically-driven sea ice volume anomalies. This physical interpretability is absent from the non-conserving state-to-state model.

While FloeNet generates accurate sea ice variability, it exhibits positive Arctic sea ice thickness biases in piControl and 1\% CO\textsubscript{2} rollouts. One might simply explain this by virtue of FloeNet being trained on a different sea ice thickness mean state to the CM4 piControl and 1\% CO\textsubscript{2} experiments. However, if FloeNet was over-fitting, then we would expect to produce thinner ice than the piControl simulation and thicker ice than the 1\% CO\textsubscript{2} simulation. The fact that FloeNet produces thicker ice with the correct spatial pattern under these different forcings is promising and suggests that the model can be improved. One current limitation is that FloeNet does not conserve energy. This could lead to ice which is too thick if FloeNet has, for example, implicitly learned to grow more ice than is actually possible given the basal heat flux forcing. In future developments, we will enforce physical energy constraints in accordance with formulations used in large-scale sea ice models \cite{Bitz1999,Collins2006}. This will involve emulating the internal ice temperature in order to close the sea ice enthalpy budget; see equations (2--5) of supplementary section S1.

Finally, subsequent work will involve coupling FloeNet to other climate model emulators. FloeNet's 6-hour timestep makes for straightforward coupling with atmospheric emulators ACE2 \cite{Watt2025} and CAMulator \cite{Chapman2025b}, both of which have a 6-hour timestep. CAMulator is trained to emulate the atmospheric component of CESM \cite{Danabasoglu2020}. FloeNet could therefore be trained to emulate the sea ice component of CESM, CICE, and then coupled to CAMulator as an interactive ice-atmosphere model. All of FloeNet's prognostic sea ice variables except the snow mass budget terms are available in CICE. Meanwhile, CAMulator predicts net surface radiative fluxes rather than downward, so would need to be modified in this respect.  

FloeNet has been built using the Ai2 emulator codebase and so can be readily coupled to ACE2 for coupled ice-atmosphere rollouts, Samudra for coupled ice-ocean rollouts, and ultimately added as a new interactive sea ice component to the SamudrACE coupled emulator \cite{Duncan2025}. We expect that this will lead to improvements in SamudrACE rollouts in a few key ways. For one, SamudrACE emulates sea ice prognostically within the ocean component using a 5-day timestep. In supplementary section S2 we show that a 5-day timestep yields poor representation of sea ice variability and trends. Moreover, a 5-day timestep has implications for accurately capturing ice surface temperature mean state and variability, as described in section \ref{sect:fluxes}. Second, SamudrACE currently has no constraint on the ocean salinity budget. We therefore anticipate that FloeNet's ice-to-ocean salt-flux predictions could be combined with precipitation fluxes from ACE2 to constrain the salt budget and, by extension, the large-scale ocean circulation. This may, in turn, lead to improved overturning circulation patterns, such as the Atlantic Meridional Overturning Circulation \cite{Sevellec2017}.

\section*{Open Research Section}
FloeNet was developed using the Ai2 climate modeling codebase, which is fully open-source (https://github.com/ai2cm/ace). The optimized weights for FloeNet and initial conditions for inference with OM4 and CM4 forcing are stored within a public repository (https://huggingface.co/M2LInES/FloeNet-OM4). Upon acceptance of this manuscript, a corresponding DOI will be created to mark FloeNet version 1.0.

\section*{Conflict of Interest declaration}
The authors declare there are no conflicts of interest for this manuscript.

\acknowledgments
William Gregory, Adam Subel, Alistair Adcroft and Laure Zanna received support through Schmidt Sciences, under the M$^2$LInES project. This work was also supported through the provisions of computational resources from the National Oceanic and Atmospheric Administration (NOAA) Geophysical Fluid Dynamics Laboratory (GFDL). Ai2 is supported by the estate of Paul G. Allen. The
authors also thank Bosong Zhang and  Danni Du for their feedback on this work.

\bibliography{agusample}

\begin{thebibliography}{}

\bibitem [\protect \citeauthoryear {%
Adcroft%
\ \protect \BOthers {.}}{%
Adcroft%
\ \protect \BOthers {.}}{%
{\protect \APACyear {2019}}%
}]{%
Adcroft2019}
\APACinsertmetastar {%
Adcroft2019}%
\begin{APACrefauthors}%
Adcroft, A.%
, Anderson, W.%
, Balaji, V.%
, Blanton, C.%
, Bushuk, M.%
, Dufour, C\BPBI O.%
\BDBL {}Zhang, R.%
\end{APACrefauthors}%
\unskip\
\newblock
\APACrefYearMonthDay{2019}{}{}.
\newblock
{\BBOQ}\APACrefatitle {The {GFDL} global ocean and sea ice model {OM4.0}: Model
  description and simulation features} {The {GFDL} global ocean and sea ice
  model {OM4.0}: Model description and simulation features}.{\BBCQ}
\newblock
\APACjournalVolNumPages{Journal of Advances in Modeling Earth
  Systems}{11}{10}{3167--3211}.
\PrintBackRefs{\CurrentBib}

\bibitem [\protect \citeauthoryear {%
Allen%
\ \protect \BOthers {.}}{%
Allen%
\ \protect \BOthers {.}}{%
{\protect \APACyear {2025}}%
}]{%
Allen2025}
\APACinsertmetastar {%
Allen2025}%
\begin{APACrefauthors}%
Allen, A.%
, Markou, S.%
, Tebbutt, W.%
, Requeima, J.%
, Bruinsma, W\BPBI P.%
, Andersson, T\BPBI R.%
\BDBL {}Turner, R\BPBI E.%
\end{APACrefauthors}%
\unskip\
\newblock
\APACrefYearMonthDay{2025}{}{}.
\newblock
{\BBOQ}\APACrefatitle {End-to-end data-driven weather prediction} {End-to-end
  data-driven weather prediction}.{\BBCQ}
\newblock
\APACjournalVolNumPages{Nature}{641}{8065}{1172--1179}.
\PrintBackRefs{\CurrentBib}

\bibitem [\protect \citeauthoryear {%
Andersson%
\ \protect \BOthers {.}}{%
Andersson%
\ \protect \BOthers {.}}{%
{\protect \APACyear {2023}}%
}]{%
Andersson2023}
\APACinsertmetastar {%
Andersson2023}%
\begin{APACrefauthors}%
Andersson, T\BPBI R.%
, Bruinsma, W\BPBI P.%
, Markou, S.%
, Requeima, J.%
, Coca-Castro, A.%
, Vaughan, A.%
\BDBL {}others%
\end{APACrefauthors}%
\unskip\
\newblock
\APACrefYearMonthDay{2023}{}{}.
\newblock
{\BBOQ}\APACrefatitle {Environmental sensor placement with convolutional
  Gaussian neural processes} {Environmental sensor placement with convolutional
  gaussian neural processes}.{\BBCQ}
\newblock
\APACjournalVolNumPages{Environmental Data Science}{2}{}{e32}.
\PrintBackRefs{\CurrentBib}

\bibitem [\protect \citeauthoryear {%
Bastin%
\ \protect \BOthers {.}}{%
Bastin%
\ \protect \BOthers {.}}{%
{\protect \APACyear {2019}}%
}]{%
Bastin2019}
\APACinsertmetastar {%
Bastin2019}%
\begin{APACrefauthors}%
Bastin, J\BHBI F.%
, Finegold, Y.%
, Garcia, C.%
, Mollicone, D.%
, Rezende, M.%
, Routh, D.%
\BDBL {}Crowther, T\BPBI W.%
\end{APACrefauthors}%
\unskip\
\newblock
\APACrefYearMonthDay{2019}{}{}.
\newblock
{\BBOQ}\APACrefatitle {The global tree restoration potential} {The global tree
  restoration potential}.{\BBCQ}
\newblock
\APACjournalVolNumPages{Science}{365}{6448}{76--79}.
\PrintBackRefs{\CurrentBib}

\bibitem [\protect \citeauthoryear {%
Behrens%
\ \protect \BOthers {.}}{%
Behrens%
\ \protect \BOthers {.}}{%
{\protect \APACyear {2025}}%
}]{%
Behrens2025}
\APACinsertmetastar {%
Behrens2025}%
\begin{APACrefauthors}%
Behrens, G.%
, Beucler, T.%
, Iglesias-Suarez, F.%
, Yu, S.%
, Gentine, P.%
, Pritchard, M.%
\BDBL {}Eyring, V.%
\end{APACrefauthors}%
\unskip\
\newblock
\APACrefYearMonthDay{2025}{}{}.
\newblock
{\BBOQ}\APACrefatitle {Simulating atmospheric processes in {E}arth system
  models and quantifying uncertainties with deep learning multi-member and
  stochastic parameterizations} {Simulating atmospheric processes in {E}arth
  system models and quantifying uncertainties with deep learning multi-member
  and stochastic parameterizations}.{\BBCQ}
\newblock
\APACjournalVolNumPages{Journal of Advances in Modeling Earth
  Systems}{17}{4}{e2024MS004272}.
\PrintBackRefs{\CurrentBib}

\bibitem [\protect \citeauthoryear {%
Bitz%
\ \BBA {} Lipscomb%
}{%
Bitz%
\ \BBA {} Lipscomb%
}{%
{\protect \APACyear {1999}}%
}]{%
Bitz1999}
\APACinsertmetastar {%
Bitz1999}%
\begin{APACrefauthors}%
Bitz, C\BPBI M.%
\BCBT {}\ \BBA {} Lipscomb, W\BPBI H.%
\end{APACrefauthors}%
\unskip\
\newblock
\APACrefYearMonthDay{1999}{}{}.
\newblock
{\BBOQ}\APACrefatitle {An energy-conserving thermodynamic model of sea ice} {An
  energy-conserving thermodynamic model of sea ice}.{\BBCQ}
\newblock
\APACjournalVolNumPages{Journal of Geophysical Research:
  Oceans}{104}{C7}{15669--15677}.
\PrintBackRefs{\CurrentBib}

\bibitem [\protect \citeauthoryear {%
Brenowitz%
\ \protect \BOthers {.}}{%
Brenowitz%
\ \protect \BOthers {.}}{%
{\protect \APACyear {2025}}%
}]{%
Brenowitz2025}
\APACinsertmetastar {%
Brenowitz2025}%
\begin{APACrefauthors}%
Brenowitz, N\BPBI D.%
, Ge, T.%
, Subramaniam, A.%
, Manshausen, P.%
, Gupta, A.%
, Hall, D\BPBI M.%
\BDBL {}Pritchard, M\BPBI S.%
\end{APACrefauthors}%
\unskip\
\newblock
\APACrefYearMonthDay{2025}{}{}.
\newblock
{\BBOQ}\APACrefatitle {Climate in a bottle: Towards a generative foundation
  model for the kilometer-scale global atmosphere} {Climate in a bottle:
  Towards a generative foundation model for the kilometer-scale global
  atmosphere}.{\BBCQ}
\newblock
\APACjournalVolNumPages{arXiv preprint arXiv:2505.06474}{}{}{}.
\PrintBackRefs{\CurrentBib}

\bibitem [\protect \citeauthoryear {%
Chapman%
\ \BBA {} Berner%
}{%
Chapman%
\ \BBA {} Berner%
}{%
{\protect \APACyear {2025}}%
}]{%
Chapman2025a}
\APACinsertmetastar {%
Chapman2025a}%
\begin{APACrefauthors}%
Chapman, W\BPBI E.%
\BCBT {}\ \BBA {} Berner, J.%
\end{APACrefauthors}%
\unskip\
\newblock
\APACrefYearMonthDay{2025}{}{}.
\newblock
{\BBOQ}\APACrefatitle {Improving climate bias and variability via CNN-based
  state-dependent model-error corrections} {Improving climate bias and
  variability via cnn-based state-dependent model-error corrections}.{\BBCQ}
\newblock
\APACjournalVolNumPages{Geophysical Research Letters}{52}{6}{e2024GL114106}.
\PrintBackRefs{\CurrentBib}

\bibitem [\protect \citeauthoryear {%
Chapman%
\ \protect \BOthers {.}}{%
Chapman%
\ \protect \BOthers {.}}{%
{\protect \APACyear {2025}}%
}]{%
Chapman2025b}
\APACinsertmetastar {%
Chapman2025b}%
\begin{APACrefauthors}%
Chapman, W\BPBI E.%
, Schreck, J\BPBI S.%
, Sha, Y.%
, Gagne~II, D\BPBI J.%
, Kimpara, D.%
, Zanna, L.%
\BDBL {}Berner, J.%
\end{APACrefauthors}%
\unskip\
\newblock
\APACrefYearMonthDay{2025}{}{}.
\newblock
{\BBOQ}\APACrefatitle {{CAM}ulator: Fast emulation of the community atmosphere
  model} {{CAM}ulator: Fast emulation of the community atmosphere
  model}.{\BBCQ}
\newblock
\APACjournalVolNumPages{arXiv preprint arXiv:2504.06007}{}{}{}.
\PrintBackRefs{\CurrentBib}

\bibitem [\protect \citeauthoryear {%
Chen%
, Mahmood%
, Tsamados%
\BCBL {}\ \BBA {} Takao%
}{%
Chen%
\ \protect \BOthers {.}}{%
{\protect \APACyear {2025}}%
}]{%
Chen2025}
\APACinsertmetastar {%
Chen2025}%
\begin{APACrefauthors}%
Chen, W.%
, Mahmood, A.%
, Tsamados, M.%
\BCBL {}\ \BBA {} Takao, S.%
\end{APACrefauthors}%
\unskip\
\newblock
\APACrefYearMonthDay{2025}{}{}.
\newblock
{\BBOQ}\APACrefatitle {Deep random features for scalable interpolation of
  spatiotemporal data} {Deep random features for scalable interpolation of
  spatiotemporal data}.{\BBCQ}
\newblock
\BIn{} \APACrefbtitle {The Thirteenth International Conference on Learning
  Representations.} {The thirteenth international conference on learning
  representations.}
\newblock
\begin{APACrefURL} \url{https://openreview.net/forum?id=OD1MV7vf41}
  \end{APACrefURL}
\PrintBackRefs{\CurrentBib}

\bibitem [\protect \citeauthoryear {%
Clark%
\ \protect \BOthers {.}}{%
Clark%
\ \protect \BOthers {.}}{%
{\protect \APACyear {2025}}%
}]{%
Clark2025}
\APACinsertmetastar {%
Clark2025}%
\begin{APACrefauthors}%
Clark, S\BPBI K.%
, Watt-Meyer, O.%
, Kwa, A.%
, McGibbon, J.%
, Henn, B.%
, Perkins, W\BPBI A.%
\BDBL {}Bretherton, C\BPBI S.%
\end{APACrefauthors}%
\unskip\
\newblock
\APACrefYearMonthDay{2025}{}{}.
\newblock
{\BBOQ}\APACrefatitle {{ACE2-SOM}: Coupling an {ML} atmospheric emulator to a
  slab ocean and learning the sensitivity of climate to changed {CO2}}
  {{ACE2-SOM}: Coupling an {ML} atmospheric emulator to a slab ocean and
  learning the sensitivity of climate to changed {CO2}}.{\BBCQ}
\newblock
\APACjournalVolNumPages{Journal of Geophysical Research: Machine Learning and
  Computation}{2}{4}{e2024JH000575}.
\PrintBackRefs{\CurrentBib}

\bibitem [\protect \citeauthoryear {%
Collins%
\ \protect \BOthers {.}}{%
Collins%
\ \protect \BOthers {.}}{%
{\protect \APACyear {2006}}%
}]{%
Collins2006}
\APACinsertmetastar {%
Collins2006}%
\begin{APACrefauthors}%
Collins, W\BPBI D.%
, Bitz, C\BPBI M.%
, Blackmon, M\BPBI L.%
, Bonan, G\BPBI B.%
, Bretherton, C\BPBI S.%
, Carton, J\BPBI A.%
\BDBL {}Smith, R\BPBI D.%
\end{APACrefauthors}%
\unskip\
\newblock
\APACrefYearMonthDay{2006}{}{}.
\newblock
{\BBOQ}\APACrefatitle {The community climate system model version 3 ({CCSM3})}
  {The community climate system model version 3 ({CCSM3})}.{\BBCQ}
\newblock
\APACjournalVolNumPages{Journal of climate}{19}{11}{2122--2143}.
\PrintBackRefs{\CurrentBib}

\bibitem [\protect \citeauthoryear {%
Danabasoglu%
\ \protect \BOthers {.}}{%
Danabasoglu%
\ \protect \BOthers {.}}{%
{\protect \APACyear {2020}}%
}]{%
Danabasoglu2020}
\APACinsertmetastar {%
Danabasoglu2020}%
\begin{APACrefauthors}%
Danabasoglu, G.%
, Lamarque, J\BHBI F.%
, Bacmeister, J.%
, Bailey, D.%
, DuVivier, A.%
, Edwards, J.%
\BDBL {}Strand, W.%
\end{APACrefauthors}%
\unskip\
\newblock
\APACrefYearMonthDay{2020}{}{}.
\newblock
{\BBOQ}\APACrefatitle {The community earth system model version 2 ({CESM}2)}
  {The community earth system model version 2 ({CESM}2)}.{\BBCQ}
\newblock
\APACjournalVolNumPages{Journal of Advances in Modeling Earth
  Systems}{12}{2}{e2019MS001916}.
\PrintBackRefs{\CurrentBib}

\bibitem [\protect \citeauthoryear {%
Dheeshjith%
\ \protect \BOthers {.}}{%
Dheeshjith%
\ \protect \BOthers {.}}{%
{\protect \APACyear {2025}}%
}]{%
Dheeshjith2025}
\APACinsertmetastar {%
Dheeshjith2025}%
\begin{APACrefauthors}%
Dheeshjith, S.%
, Subel, A.%
, Adcroft, A.%
, Busecke, J.%
, Fernandez-Granda, C.%
, Gupta, S.%
\BCBL {}\ \BBA {} Zanna, L.%
\end{APACrefauthors}%
\unskip\
\newblock
\APACrefYearMonthDay{2025}{}{}.
\newblock
{\BBOQ}\APACrefatitle {Samudra: An {AI} global ocean emulator for climate}
  {Samudra: An {AI} global ocean emulator for climate}.{\BBCQ}
\newblock
\APACjournalVolNumPages{Geophysical Research Letters}{52}{10}{e2024GL114318}.
\PrintBackRefs{\CurrentBib}

\bibitem [\protect \citeauthoryear {%
Duncan%
\ \protect \BOthers {.}}{%
Duncan%
\ \protect \BOthers {.}}{%
{\protect \APACyear {2025}}%
}]{%
Duncan2025}
\APACinsertmetastar {%
Duncan2025}%
\begin{APACrefauthors}%
Duncan, J\BPBI P.%
, Wu, E.%
, Dheeshjith, S.%
, Subel, A.%
, Arcomano, T.%
, Clark, S\BPBI K.%
\BDBL {}Bretherton, C.%
\end{APACrefauthors}%
\unskip\
\newblock
\APACrefYearMonthDay{2025}{}{}.
\newblock
{\BBOQ}\APACrefatitle {Samudr{ACE}: Fast and accurate coupled climate modeling
  with {3D} ocean and atmosphere emulators} {Samudr{ACE}: Fast and accurate
  coupled climate modeling with {3D} ocean and atmosphere emulators}.{\BBCQ}
\newblock
\APACjournalVolNumPages{arXiv preprint arXiv:2509.12490}{}{}{}.
\PrintBackRefs{\CurrentBib}

\bibitem [\protect \citeauthoryear {%
Durand%
\ \protect \BOthers {.}}{%
Durand%
\ \protect \BOthers {.}}{%
{\protect \APACyear {2024}}%
}]{%
Durand2024}
\APACinsertmetastar {%
Durand2024}%
\begin{APACrefauthors}%
Durand, C.%
, Finn, T\BPBI S.%
, Farchi, A.%
, Bocquet, M.%
, Boutin, G.%
\BCBL {}\ \BBA {} {\'O}lason, E.%
\end{APACrefauthors}%
\unskip\
\newblock
\APACrefYearMonthDay{2024}{}{}.
\newblock
{\BBOQ}\APACrefatitle {Data-driven surrogate modeling of high-resolution
  sea-ice thickness in the {A}rctic} {Data-driven surrogate modeling of
  high-resolution sea-ice thickness in the {A}rctic}.{\BBCQ}
\newblock
\APACjournalVolNumPages{The Cryosphere}{18}{4}{1791--1815}.
\PrintBackRefs{\CurrentBib}

\bibitem [\protect \citeauthoryear {%
Falasca%
}{%
Falasca%
}{%
{\protect \APACyear {2025}}%
}]{%
Falasca2025}
\APACinsertmetastar {%
Falasca2025}%
\begin{APACrefauthors}%
Falasca, F.%
\end{APACrefauthors}%
\unskip\
\newblock
\APACrefYearMonthDay{2025}{}{}.
\newblock
{\BBOQ}\APACrefatitle {Probing forced responses and causality in data-driven
  climate emulators: {C}onceptual limitations and the role of reduced-order
  models} {Probing forced responses and causality in data-driven climate
  emulators: {C}onceptual limitations and the role of reduced-order
  models}.{\BBCQ}
\newblock
\APACjournalVolNumPages{Physical Review Research}{7}{4}{043314}.
\PrintBackRefs{\CurrentBib}

\bibitem [\protect \citeauthoryear {%
Finn%
\ \protect \BOthers {.}}{%
Finn%
\ \protect \BOthers {.}}{%
{\protect \APACyear {2025}}%
}]{%
Finn2025}
\APACinsertmetastar {%
Finn2025}%
\begin{APACrefauthors}%
Finn, T\BPBI S.%
, Bocquet, M.%
, Rampal, P.%
, Durand, C.%
, Porro, F.%
, Farchi, A.%
\BCBL {}\ \BBA {} Carrassi, A.%
\end{APACrefauthors}%
\unskip\
\newblock
\APACrefYearMonthDay{2025}{}{}.
\newblock
{\BBOQ}\APACrefatitle {Generative {AI} models enable efficient and physically
  consistent sea-ice simulations} {Generative {AI} models enable efficient and
  physically consistent sea-ice simulations}.{\BBCQ}
\newblock
\APACjournalVolNumPages{arXiv preprint arXiv:2508.14984}{}{}{}.
\PrintBackRefs{\CurrentBib}

\bibitem [\protect \citeauthoryear {%
Gregory%
\ \protect \BOthers {.}}{%
Gregory%
\ \protect \BOthers {.}}{%
{\protect \APACyear {2026}}%
}]{%
Gregory2026}
\APACinsertmetastar {%
Gregory2026}%
\begin{APACrefauthors}%
Gregory, W.%
, Bushuk, M.%
, Zhang, Y\BHBI F.%
, Adcroft, A.%
, Zanna, L.%
, McHugh, C.%
\BCBL {}\ \BBA {} Jia, L.%
\end{APACrefauthors}%
\unskip\
\newblock
\APACrefYearMonthDay{2026}{}{}.
\newblock
{\BBOQ}\APACrefatitle {Advancing global sea ice prediction capabilities using a
  fully coupled climate model with integrated machine learning} {Advancing
  global sea ice prediction capabilities using a fully coupled climate model
  with integrated machine learning}.{\BBCQ}
\newblock
\APACjournalVolNumPages{Science Advances}{12}{1}{eady8957}.
\PrintBackRefs{\CurrentBib}

\bibitem [\protect \citeauthoryear {%
Gregory%
\ \protect \BOthers {.}}{%
Gregory%
\ \protect \BOthers {.}}{%
{\protect \APACyear {2024}}%
}]{%
Gregory2024}
\APACinsertmetastar {%
Gregory2024}%
\begin{APACrefauthors}%
Gregory, W.%
, MacEachern, R.%
, Takao, S.%
, Lawrence, I\BPBI R.%
, Nab, C.%
, Deisenroth, M\BPBI P.%
\BCBL {}\ \BBA {} Tsamados, M.%
\end{APACrefauthors}%
\unskip\
\newblock
\APACrefYearMonthDay{2024}{}{}.
\newblock
{\BBOQ}\APACrefatitle {Scalable interpolation of satellite altimetry data with
  probabilistic machine learning} {Scalable interpolation of satellite
  altimetry data with probabilistic machine learning}.{\BBCQ}
\newblock
\APACjournalVolNumPages{Nature Communications}{15}{1}{7453}.
\PrintBackRefs{\CurrentBib}

\bibitem [\protect \citeauthoryear {%
Hahn%
, Armour%
, Zelinka%
, Bitz%
\BCBL {}\ \BBA {} Donohoe%
}{%
Hahn%
\ \protect \BOthers {.}}{%
{\protect \APACyear {2021}}%
}]{%
Hahn2021}
\APACinsertmetastar {%
Hahn2021}%
\begin{APACrefauthors}%
Hahn, L\BPBI C.%
, Armour, K\BPBI C.%
, Zelinka, M\BPBI D.%
, Bitz, C\BPBI M.%
\BCBL {}\ \BBA {} Donohoe, A.%
\end{APACrefauthors}%
\unskip\
\newblock
\APACrefYearMonthDay{2021}{}{}.
\newblock
{\BBOQ}\APACrefatitle {Contributions to polar amplification in {CMIP}5 and
  {CMIP}6 models} {Contributions to polar amplification in {CMIP}5 and {CMIP}6
  models}.{\BBCQ}
\newblock
\APACjournalVolNumPages{Frontiers in Earth Science}{9}{}{710036}.
\PrintBackRefs{\CurrentBib}

\bibitem [\protect \citeauthoryear {%
Held%
\ \protect \BOthers {.}}{%
Held%
\ \protect \BOthers {.}}{%
{\protect \APACyear {2019}}%
}]{%
Held2019}
\APACinsertmetastar {%
Held2019}%
\begin{APACrefauthors}%
Held, I\BPBI M.%
, Guo, H.%
, Adcroft, A.%
, Dunne, J\BPBI P.%
, Horowitz, L\BPBI W.%
, Krasting, J\BPBI P.%
\BDBL {}Zadeh, N.%
\end{APACrefauthors}%
\unskip\
\newblock
\APACrefYearMonthDay{2019}{}{}.
\newblock
{\BBOQ}\APACrefatitle {Structure and performance of {GFDL}'s {CM4}.0 climate
  model} {Structure and performance of {GFDL}'s {CM4}.0 climate model}.{\BBCQ}
\newblock
\APACjournalVolNumPages{Journal of Advances in Modeling Earth
  Systems}{11}{11}{3691--3727}.
\PrintBackRefs{\CurrentBib}

\bibitem [\protect \citeauthoryear {%
Keen%
\ \protect \BOthers {.}}{%
Keen%
\ \protect \BOthers {.}}{%
{\protect \APACyear {2021}}%
}]{%
Keen2021}
\APACinsertmetastar {%
Keen2021}%
\begin{APACrefauthors}%
Keen, A.%
, Blockley, E.%
, Bailey, D\BPBI A.%
, Boldingh~Debernard, J.%
, Bushuk, M.%
, Delhaye, S.%
\BDBL {}Vancoppenolle, M.%
\end{APACrefauthors}%
\unskip\
\newblock
\APACrefYearMonthDay{2021}{}{}.
\newblock
{\BBOQ}\APACrefatitle {An inter-comparison of the mass budget of the {A}rctic
  sea ice in {CMIP}6 models} {An inter-comparison of the mass budget of the
  {A}rctic sea ice in {CMIP}6 models}.{\BBCQ}
\newblock
\APACjournalVolNumPages{The Cryosphere}{15}{2}{951--982}.
\PrintBackRefs{\CurrentBib}

\bibitem [\protect \citeauthoryear {%
Kochkov%
\ \protect \BOthers {.}}{%
Kochkov%
\ \protect \BOthers {.}}{%
{\protect \APACyear {2024}}%
}]{%
Kochkov2024}
\APACinsertmetastar {%
Kochkov2024}%
\begin{APACrefauthors}%
Kochkov, D.%
, Yuval, J.%
, Langmore, I.%
, Norgaard, P.%
, Smith, J.%
, Mooers, G.%
\BDBL {}Hoyer, S.%
\end{APACrefauthors}%
\unskip\
\newblock
\APACrefYearMonthDay{2024}{}{}.
\newblock
{\BBOQ}\APACrefatitle {Neural general circulation models for weather and
  climate} {Neural general circulation models for weather and climate}.{\BBCQ}
\newblock
\APACjournalVolNumPages{Nature}{632}{8027}{1060--1066}.
\PrintBackRefs{\CurrentBib}

\bibitem [\protect \citeauthoryear {%
Lam%
\ \protect \BOthers {.}}{%
Lam%
\ \protect \BOthers {.}}{%
{\protect \APACyear {2023}}%
}]{%
Lam2023}
\APACinsertmetastar {%
Lam2023}%
\begin{APACrefauthors}%
Lam, R.%
, Sanchez-Gonzalez, A.%
, Willson, M.%
, Wirnsberger, P.%
, Fortunato, M.%
, Alet, F.%
\BDBL {}Battaglia, P.%
\end{APACrefauthors}%
\unskip\
\newblock
\APACrefYearMonthDay{2023}{}{}.
\newblock
{\BBOQ}\APACrefatitle {Learning skillful medium-range global weather
  forecasting} {Learning skillful medium-range global weather
  forecasting}.{\BBCQ}
\newblock
\APACjournalVolNumPages{Science}{382}{6677}{1416--1421}.
\PrintBackRefs{\CurrentBib}

\bibitem [\protect \citeauthoryear {%
Mansfield%
\ \BBA {} Sheshadri%
}{%
Mansfield%
\ \BBA {} Sheshadri%
}{%
{\protect \APACyear {2024}}%
}]{%
Mansfield2024}
\APACinsertmetastar {%
Mansfield2024}%
\begin{APACrefauthors}%
Mansfield, L\BPBI A.%
\BCBT {}\ \BBA {} Sheshadri, A.%
\end{APACrefauthors}%
\unskip\
\newblock
\APACrefYearMonthDay{2024}{}{}.
\newblock
{\BBOQ}\APACrefatitle {Uncertainty quantification of a machine learning
  subgrid-scale parameterization for atmospheric gravity waves} {Uncertainty
  quantification of a machine learning subgrid-scale parameterization for
  atmospheric gravity waves}.{\BBCQ}
\newblock
\APACjournalVolNumPages{Journal of Advances in Modeling Earth
  Systems}{16}{7}{e2024MS004292}.
\PrintBackRefs{\CurrentBib}

\bibitem [\protect \citeauthoryear {%
Mardani%
\ \protect \BOthers {.}}{%
Mardani%
\ \protect \BOthers {.}}{%
{\protect \APACyear {2025}}%
}]{%
Mardani2025}
\APACinsertmetastar {%
Mardani2025}%
\begin{APACrefauthors}%
Mardani, M.%
, Brenowitz, N.%
, Cohen, Y.%
, Pathak, J.%
, Chen, C\BHBI Y.%
, Liu, C\BHBI C.%
\BDBL {}Pritchard, M.%
\end{APACrefauthors}%
\unskip\
\newblock
\APACrefYearMonthDay{2025}{}{}.
\newblock
{\BBOQ}\APACrefatitle {Residual corrective diffusion modeling for km-scale
  atmospheric downscaling} {Residual corrective diffusion modeling for km-scale
  atmospheric downscaling}.{\BBCQ}
\newblock
\APACjournalVolNumPages{Communications Earth \& Environment}{6}{1}{124}.
\PrintBackRefs{\CurrentBib}

\bibitem [\protect \citeauthoryear {%
Notz%
\ \BBA {} Stroeve%
}{%
Notz%
\ \BBA {} Stroeve%
}{%
{\protect \APACyear {2016}}%
}]{%
Notz2016}
\APACinsertmetastar {%
Notz2016}%
\begin{APACrefauthors}%
Notz, D.%
\BCBT {}\ \BBA {} Stroeve, J.%
\end{APACrefauthors}%
\unskip\
\newblock
\APACrefYearMonthDay{2016}{}{}.
\newblock
{\BBOQ}\APACrefatitle {Observed {A}rctic sea-ice loss directly follows
  anthropogenic {CO2} emission} {Observed {A}rctic sea-ice loss directly
  follows anthropogenic {CO2} emission}.{\BBCQ}
\newblock
\APACjournalVolNumPages{Science}{354}{6313}{747--750}.
\PrintBackRefs{\CurrentBib}

\bibitem [\protect \citeauthoryear {%
Paciorek%
\ \BBA {} Cooley%
}{%
Paciorek%
\ \BBA {} Cooley%
}{%
{\protect \APACyear {2025}}%
}]{%
Paciorek2025}
\APACinsertmetastar {%
Paciorek2025}%
\begin{APACrefauthors}%
Paciorek, C\BPBI J.%
\BCBT {}\ \BBA {} Cooley, D.%
\end{APACrefauthors}%
\unskip\
\newblock
\APACrefYearMonthDay{2025}{}{}.
\newblock
{\BBOQ}\APACrefatitle {Quantifying very extreme precipitation and temperature
  using huge ensembles generated by machine learning-based climate model
  emulators} {Quantifying very extreme precipitation and temperature using huge
  ensembles generated by machine learning-based climate model
  emulators}.{\BBCQ}
\newblock
\APACjournalVolNumPages{arXiv preprint arXiv:2510.08893}{}{}{}.
\PrintBackRefs{\CurrentBib}

\bibitem [\protect \citeauthoryear {%
Pathak%
\ \protect \BOthers {.}}{%
Pathak%
\ \protect \BOthers {.}}{%
{\protect \APACyear {2022}}%
}]{%
Pathak2022}
\APACinsertmetastar {%
Pathak2022}%
\begin{APACrefauthors}%
Pathak, J.%
, Subramanian, S.%
, Harrington, P.%
, Raja, S.%
, Chattopadhyay, A.%
, Mardani, M.%
\BDBL {}Anandkumar, A.%
\end{APACrefauthors}%
\unskip\
\newblock
\APACrefYearMonthDay{2022}{}{}.
\newblock
{\BBOQ}\APACrefatitle {Fourcastnet: {A} global data-driven high-resolution
  weather model using adaptive fourier neural operators} {Fourcastnet: {A}
  global data-driven high-resolution weather model using adaptive fourier
  neural operators}.{\BBCQ}
\newblock
\APACjournalVolNumPages{arXiv preprint arXiv:2202.11214}{}{}{}.
\PrintBackRefs{\CurrentBib}

\bibitem [\protect \citeauthoryear {%
Perkins%
\ \protect \BOthers {.}}{%
Perkins%
\ \protect \BOthers {.}}{%
{\protect \APACyear {2025}}%
}]{%
Perkins2025}
\APACinsertmetastar {%
Perkins2025}%
\begin{APACrefauthors}%
Perkins, W\BPBI A.%
, Kwa, A.%
, McGibbon, J.%
, Arcomano, T.%
, Clark, S\BPBI K.%
, Watt-Meyer, O.%
\BDBL {}Harris, L\BPBI M.%
\end{APACrefauthors}%
\unskip\
\newblock
\APACrefYearMonthDay{2025}{}{}.
\newblock
{\BBOQ}\APACrefatitle {{HiRO-ACE}: Fast and skillful {AI} emulation and
  downscaling trained on a 3 km global storm-resolving model} {{HiRO-ACE}: Fast
  and skillful {AI} emulation and downscaling trained on a 3 km global
  storm-resolving model}.{\BBCQ}
\newblock
\APACjournalVolNumPages{arXiv preprint arXiv:2512.18224}{}{}{}.
\PrintBackRefs{\CurrentBib}

\bibitem [\protect \citeauthoryear {%
S{\'e}vellec%
, Fedorov%
\BCBL {}\ \BBA {} Liu%
}{%
S{\'e}vellec%
\ \protect \BOthers {.}}{%
{\protect \APACyear {2017}}%
}]{%
Sevellec2017}
\APACinsertmetastar {%
Sevellec2017}%
\begin{APACrefauthors}%
S{\'e}vellec, F.%
, Fedorov, A\BPBI V.%
\BCBL {}\ \BBA {} Liu, W.%
\end{APACrefauthors}%
\unskip\
\newblock
\APACrefYearMonthDay{2017}{}{}.
\newblock
{\BBOQ}\APACrefatitle {Arctic sea-ice decline weakens the {A}tlantic meridional
  overturning circulation} {Arctic sea-ice decline weakens the {A}tlantic
  meridional overturning circulation}.{\BBCQ}
\newblock
\APACjournalVolNumPages{Nature Climate Change}{7}{8}{604--610}.
\PrintBackRefs{\CurrentBib}

\bibitem [\protect \citeauthoryear {%
Tsujino%
\ \protect \BOthers {.}}{%
Tsujino%
\ \protect \BOthers {.}}{%
{\protect \APACyear {2018}}%
}]{%
Tsujino2018}
\APACinsertmetastar {%
Tsujino2018}%
\begin{APACrefauthors}%
Tsujino, H.%
, Urakawa, S.%
, Nakano, H.%
, Small, R\BPBI J.%
, Kim, W\BPBI M.%
, Yeager, S\BPBI G.%
\BDBL {}Yamazaki, D.%
\end{APACrefauthors}%
\unskip\
\newblock
\APACrefYearMonthDay{2018}{}{}.
\newblock
{\BBOQ}\APACrefatitle {{JRA-55} based surface dataset for driving
  ocean--sea-ice models ({JRA55}-do)} {{JRA-55} based surface dataset for
  driving ocean--sea-ice models ({JRA55}-do)}.{\BBCQ}
\newblock
\APACjournalVolNumPages{Ocean Modelling}{130}{}{79--139}.
\PrintBackRefs{\CurrentBib}

\bibitem [\protect \citeauthoryear {%
Watt-Meyer%
\ \protect \BOthers {.}}{%
Watt-Meyer%
\ \protect \BOthers {.}}{%
{\protect \APACyear {2021}}%
}]{%
Watt2021}
\APACinsertmetastar {%
Watt2021}%
\begin{APACrefauthors}%
Watt-Meyer, O.%
, Brenowitz, N\BPBI D.%
, Clark, S\BPBI K.%
, Henn, B.%
, Kwa, A.%
, McGibbon, J.%
\BDBL {}Bretherton, C\BPBI S.%
\end{APACrefauthors}%
\unskip\
\newblock
\APACrefYearMonthDay{2021}{}{}.
\newblock
{\BBOQ}\APACrefatitle {Correcting weather and climate models by machine
  learning nudged historical simulations} {Correcting weather and climate
  models by machine learning nudged historical simulations}.{\BBCQ}
\newblock
\APACjournalVolNumPages{Geophysical Research Letters}{48}{15}{e2021GL092555}.
\PrintBackRefs{\CurrentBib}

\bibitem [\protect \citeauthoryear {%
Watt-Meyer%
\ \protect \BOthers {.}}{%
Watt-Meyer%
\ \protect \BOthers {.}}{%
{\protect \APACyear {2025}}%
}]{%
Watt2025}
\APACinsertmetastar {%
Watt2025}%
\begin{APACrefauthors}%
Watt-Meyer, O.%
, Henn, B.%
, McGibbon, J.%
, Clark, S\BPBI K.%
, Kwa, A.%
, Perkins, W\BPBI A.%
\BDBL {}Bretherton, C\BPBI S.%
\end{APACrefauthors}%
\unskip\
\newblock
\APACrefYearMonthDay{2025}{}{}.
\newblock
{\BBOQ}\APACrefatitle {{ACE}2: accurately learning subseasonal to decadal
  atmospheric variability and forced responses} {{ACE}2: accurately learning
  subseasonal to decadal atmospheric variability and forced responses}.{\BBCQ}
\newblock
\APACjournalVolNumPages{npj Climate and Atmospheric Science}{8}{1}{205}.
\PrintBackRefs{\CurrentBib}

\bibitem [\protect \citeauthoryear {%
Zanna%
\ \protect \BOthers {.}}{%
Zanna%
\ \protect \BOthers {.}}{%
{\protect \APACyear {2025}}%
}]{%
Zanna2025}
\APACinsertmetastar {%
Zanna2025}%
\begin{APACrefauthors}%
Zanna, L.%
, Gregory, W.%
, Perezhogin, P.%
, Sane, A.%
, Zhang, C.%
, Adcroft, A.%
\BDBL {}Wu, J.%
\end{APACrefauthors}%
\unskip\
\newblock
\APACrefYearMonthDay{2025}{}{}.
\newblock
{\BBOQ}\APACrefatitle {A framework for hybrid physics-{AI} coupled ocean
  models} {A framework for hybrid physics-{AI} coupled ocean models}.{\BBCQ}
\newblock
\APACjournalVolNumPages{arXiv preprint arXiv:2510.22676}{}{}{}.
\PrintBackRefs{\CurrentBib}

\end{thebibliography}
\newpage

\renewcommand{\thefigure}{S\arabic{figure}}
\renewcommand{\thetable}{S\arabic{table}}
\renewcommand{\theequation}{S\arabic{equation}}
\renewcommand{\thepage}{S\arabic{page}}
\setcounter{figure}{0}
\setcounter{section}{0}
\setcounter{table}{0}
\setcounter{equation}{0}
\setcounter{page}{1}

\title{Supporting Information for ``FloeNet: A mass-conserving global sea ice emulator that generalizes across climates''}

\authors{William Gregory\affil{1}, Mitchell Bushuk\affil{2}, James Duncan\affil{3}, Elynn Wu\affil{3}, Adam Subel\affil{4}, Spencer K. Clark\affil{2,3}, Bill Hurlin\affil{2}, Oliver Watt-Meyer\affil{3}, Alistair Adcroft\affil{1}, Chris Bretherton\affil{3}, Laure Zanna\affil{4}}

\affiliation{1}{Atmospheric and Oceanic Sciences Program, Princeton University, NJ, USA}
\affiliation{2}{Geophysical Fluid Dynamics Laboratory, NOAA, Princeton, NJ, USA}
\affiliation{3}{Allen Institute for Artificial Intelligence (Ai2), Seattle, USA}
\affiliation{4}{Courant Institute School of Mathematics, Computing, and Data Science, New York University, New York, NY, USA}
\vspace{10pt}
\noindent\textbf{Contents of this file}
\begin{enumerate}
\item Supplementary Methods: FloeNet architecture
\item Supplementary Results: Sensitivity tests
\item Tables S1 and S2
\item Figures S1 to S18
\end{enumerate}
\vspace{10pt}

\noindent\textbf{Introduction}
In this supporting information we provide further details on the decisions behind various aspects of the FloeNet architecture and training procedure. We also provide additional results on sensitivity tests, including random seed, icosahedral mesh resolution, input forcing, timestepping, and architecture. Supplementary figures mentioned in the main text are also included as Figs. S1--S9.

\section{FloeNet architecture}

\subsection*{Graph neural network}
FloeNet is a graph neural network (GNN) model, based on the encoder-processor-decoder framework used in GraphCast (Lam et al., 2023). The encoder is a GNN which performs a learned mapping of input variables to 256 latent features on an icosahedral multi-resolution mesh (see Table \ref{tab:tab1} for a full list of FloeNet's input and output variable names). The multi-resolution mesh, or multi-mesh, is defined by iteratively refining a regular 12-node icosahedron, where each refinement divides each triangular face into four smaller triangles. With M$_0$ as the regular 12-node icosahedron, each subsequent refinement can be labelled accordingly \{M$_0$, M$_1$, ..., M$_i$\}. In FloeNet, we perform 6 refinements and then discard M$_0$ through M$_3$, leaving our final mesh as M$_4$, M$_5$, and M$_6$. The node positions of the multi-mesh are defined by M$_6$ (40,962 total nodes) and then edges can connect across the different mesh resolutions, allowing for efficient long-range communication. The encoder therefore maps local input grid cells to a given node of M$_6$, where locality is determined by a user-defined fraction of the maximum edge distance of M$_6$. We set this fraction to 1.0, which results in an average of 4 grid cells mapping to a given mesh node. Similar to GraphCast, we also embed the spatial and relative positions of each grid and mesh node into the encoder.

The processor of FloeNet consists of 4 stacked GNN layers with independent weights, which perform message-passing on the multi-mesh. The processor updates each node's features and edges using information from adjacent nodes. With each subsequent GNN layer, information becomes progressively more non-local, as nodes continue to update their local neighborhood of node features and edges. Similar to the encoder, our processor feature dimension is 256. 

The final step involves a decoder, which performs a learned mapping of features on the processor multi-mesh to output variables on the original latitude-longitude grid. Similar to GraphCast, each grid point connects to three mesh nodes by finding the triangular face in M$_6$ whose area contains that grid point. The three mesh nodes corresponding to vertices of that face then form the unique mesh-to-grid connection.

Finally, we note that all of FloeNet's internal neural networks are multi-layer perceptrons with an input layer, single hidden layer, and output layer. The hidden layer activations are all generated using a sigmoid linear unit (SiLU) activation function, and all output layers (except the final output step of the decoder) apply layer-wise normalization.

\subsection*{Mass and area budget implementation}
Our mass-conservative approach is implemented as a correction to FloeNet's prediction of \texttt{siconc}, \texttt{simass}, and \texttt{sisnmass} at each timestep. Taking \texttt{simass} as an example, we first estimate a new mass state by scaling the predicted source, sink, and transport terms (\texttt{LSRCi}, \texttt{LSNKi}, and \texttt{XPRTi}, respectively) by the 6-hour timestep, and adding each term to the mass from the previous timestep. Because \texttt{simass} must be non-negative, we then identify any grid points in the updated mass which are negative and redistribute the excess tendency across one or more budget terms, to ensure that the updated \texttt{simass} equals zero. For example, if the updated mass at a given grid point is $-3$ kgm$^{-2}$ and all three budget terms have non-zero predictions, then we add 1 kgm$^{-2}$ to \texttt{LSRCi}, \texttt{LSNKi}, and \texttt{XPRTi}. If this adjustment then results in either \texttt{LSRCi} becoming negative or \texttt{LSNKi} becoming positive, we remove the excess tendency from that term and add it to \texttt{XPRTi}. Furthermore, we note that in the example above, if \texttt{LSRCi} was initially zero (for example, in summer), then we add 1.5 kgm$^{-2}$ to \texttt{LSNKi} and \texttt{XPRTi}. If \texttt{LSRCi} and \texttt{LSNKi} were both zero, then we add 3 kgm$^{-2}$ to \texttt{XPRTi}. In summary, we do not change the net mass flux predicted by FloeNet, we simply redistribute the mass change across the budget terms to ensure physical bounds on each variable. We use the same approach to ensure that \texttt{siconc} has an upper bound of 1. Finally, to ensure closure between the budget terms and the updated state, we do not clamp \texttt{siconc} to have minimum and maximum values of 0 and 1. Similarly, we do not clamp \texttt{simass} and \texttt{sisnmass} to have minimum values of 0. Therefore, due to roundoff, we have noticed some grid cells contain negative values between $-1 \times 10^{-16}$ and $-1 \times 10^{-19}$. We therefore leave it to the discretion of the user to apply clamping as a post-processing.

It is also worth noting that, to avoid situations where \texttt{simass} is equal to zero but \texttt{siconc} or \texttt{sisnmass} are greater than zero, we use the updated \texttt{simass} as a mask when updating \texttt{siconc} and \texttt{sisnmass}. When updating \texttt{siconc}, for example, at all locations where the new \texttt{simass} is zero, we adjust the \texttt{siconc} budgets to produce an updated \texttt{siconc} equal to zero at those grid points. We recognize that this approach does not account for the case when \texttt{simass} is greater than zero but \texttt{siconc} or \texttt{sisnmass} are zero, as this would mean adding some arbitrary amount of area or mass to those grid points. We may wish to consider this for future iterations of FloeNet.

\subsection*{Physically-motivated forcing variable selection}

In the main text, we provide motivation for our choice of prognostic and diagnostic variables, emphasizing that learning area and mass budgets yields an interpretable and generalizable model, and that salt and energy fluxes could be useful when coupled to atmosphere and ocean emulators. In this section, we provide more details on our motivation for choice of forcing variables, linking each to their physical role in sea ice dynamics and thermodynamics. For dynamics, large-scale sea ice models solve some variation of the following 2-dimensional momentum equation:

\begin{equation}
m\frac{\partial\mathbf{u}}{\partial t} = \nabla\cdot\sigma + \tau_a + \tau_w - mf\hat{\mathbf{k}}\times \mathbf{u} - mg\nabla H.
\label{eq:momentum}
\end{equation}

Here, $m$ is the sea ice mass (per unit area), $\mathbf{u}$ is the sea ice velocity, $\sigma$ is the vertically-integrated internal ice stress tensor which describes how sea ice deforms under tensile, shear, and compressive stress regimes, $\tau_a$ is the wind stress over sea ice, $\tau_w$ is the ice-ocean stress, $mf\hat{\mathbf{k}}\times \mathbf{u}$ is a stress term due to the Coriolis force, and $mg\nabla H$ is a stress due to sea-surface height gradients. Given this understanding, we provide $\tau_a$ (\texttt{FA\_X}, \texttt{FA\_Y}), $\tau_w$ (\texttt{FW\_X}, \texttt{FW\_Y}), and $H$ (\texttt{SSH}) as forcing variables, to constrain the sea ice dynamics. We then allow FloeNet to learn its own representation of internal ice stresses and stress due to Coriolis. Finally, we  note that \texttt{FA\_X}, \texttt{FA\_Y}, \texttt{FW\_X}, and \texttt{FW\_Y} are the names of the SIS2 diagnostics, which are only defined over sea ice covered grid cells. However, we would like FloeNet to learn non-local ice-atmosphere and ice-ocean interactions, for example, across the open ocean and sea ice covered grid cells spanning the ice edge. We therefore utilize the equivalent ice-ocean stress diagnostics from MOM6 (\texttt{tauuo} and \texttt{tauvo}), which are defined over all ocean grid cells. In summary, for the wind stress field, we use \texttt{FA\_X} values at grid points where \texttt{siconc} is greater than or equal to 15\% and merge this with \texttt{tauuo} at ice-free gridpoints (and similarly for \texttt{tauvo} and \texttt{FA\_Y}). The ice-ocean stress is then given as \texttt{tauuo} and \texttt{tauvo}. Therefore, before normalization our wind stress fields and ice-ocean stress fields are identical where \texttt{siconc} is less than 15\%.

Following Bitz and Lipscomb (1999), we can also define the energy needed to thermodynamically raise the temperature of a unit volume of sea ice with salinity $S$ and temperature $T$, and then fully melt it:
\begin{equation}
q(S,T) = \rho_i c_i(T_m - T) + \rho_i L_i\bigg(1 - \frac{T_m}{T}\bigg).
\label{eq:enthalpy}
\end{equation}
Here, $c_i$ is the specific heat capacity of ice, $T_m = -0.054S$ is an empirical salinity-dependent melting temperature of ice, and $L_i$ is the latent heat of fusion of ice. For snow, (\ref{eq:enthalpy}) becomes  $q(T) = -\rho_s c_s T + \rho_s L_s$. This $q(\cdot)$ term is more commonly known as the enthalpy of sea ice or snow. The bottom surface flux balance for sea ice melt and growth is then given by:
\begin{equation}
F_w - k\frac{\partial T}{\partial z} = -\frac{\partial h_i}{\partial t}q(S,T).
\label{eq:botflux}
\end{equation}
Here, $F_w = \beta(T_w - T_f)$ is the ocean-to-ice heat flux (\texttt{BHEAT}), where $T_w$ is the ocean surface temperature, $T_f = -0.054S_w$ is the salinity-dependent freezing temperature of sea water, and $\beta$ is a thermodynamic coupling coefficient with a value of 240 Wm$^{-2}$K$^{-1}$. The term $k\frac{\partial T}{\partial z}$ is the conductive heat flux upwards through the sea ice at the ice-ocean interface, with ice thermal conductivity given by $k$.  The right-hand-side of (\ref{eq:botflux}) is equal to \texttt{BMELT}. To constrain the bottom surface flux balance, we therefore force FloeNet with $F_w$ (\texttt{BHEAT}). We also include the ice-surface skin temperature (\texttt{TS}) and the sea ice thickness (via \texttt{simass} and \texttt{siconc}), which strongly constrain the conductive heat flux.

The top surface energy flux balance is given by:
\begin{equation}
F(T_0) = \texttt{SW} - I_0 + \texttt{LW} + \texttt{SH} + \texttt{LH} + k\frac{\partial T}{\partial z},
\label{eq:topflux}
\end{equation}
where $T_0$ denotes the surface temperature of the sea ice or snow, $\texttt{SW}$ is the net downward shortwave radiation, $I_0$ is the amount of downward shortwave that penetrates the surface, \texttt{LW} is the net longwave radiation, with the outgoing longwave contribution defined by $\kappa T_0^4$ ($\kappa$ is the Stefan-Boltzmann constant). \texttt{SH} and \texttt{LH} are then sensible and latent heat fluxes and $k\frac{\partial T}{\partial z}$ is the conductive heat flux upward through the snow or ice. When $F(T_m)\geq 0$, the upper surface is fixed at the melting temperature for sea ice or snow and melting occurs according to:
\begin{equation}
F(T_m) = -\frac{\partial h_i}{\partial t}q(S,T_m), 
\label{eq:melting1}
\end{equation}
where the right-hand-side is now equal to \texttt{TMELT}. To constrain the top surface energy balance, we force FloeNet with downward shortwave radiation (\texttt{SWDN}), downward longwave radiation (\texttt{LWDN}), \texttt{SH} and \texttt{LH}. Note that we also force with rate of snow fall (\texttt{SNOWFL}) to constrain snow-on-sea-ice mass. Each of these fluxes are readily available from atmospheric model components. It is worth highlighting here that we force FloeNet with  \texttt{SWDN} rather than \texttt{SW}, as \texttt{SW} depends on the surface albedo ($\alpha$), given by $\texttt{SW} = \texttt{SWDN}(1 - \alpha)$. In reality, the surface albedo depends on the surface conditions of the sea ice and snow. Therefore, forcing FloeNet with implicit albedo information could lead to data leakage issues. In principle, FloeNet could prognostically emulate albedo. However, this poses challenges, as the SIS2 albedo diagnostic is not well defined when \texttt{SWDN} is zero. We note that FloeNet's predictions of \texttt{TS}, \texttt{sisnmass}, \texttt{simass}, and \texttt{siconc} allow us to analytically derive albedo estimates at each grid point, via the CCSM3 scheme (Collins et al., 2006). We do not show results of this here, as this scheme has several free parameters which require tuning, including weights for the direct and diffuse components of visible and near infrared incoming solar radiation, and the area fraction of snow cover over sea ice. We therefore leave this for future work.

\subsection*{Training details}
Prior to training, we normalize all input and output variables to zero mean and unit variance, with statistics computed over the 1969--2005 training period and over grid cells poleward of 50$^\circ$ latitude. We then rely on the Ai2 climate model emulator codebase for training and validation. This codebase allows for efficient training pipelines, with features such as distributed GPU training, activation checkpointing, and gradient accumulation (detaching gradients from the computational graph). When training FloeNet, we utilize the distributed training functionality to train across 4 NVIDIA H200 GPUs. With a batch size of 32, each GPU therefore processes 8 samples. We also use gradient accumulation for memory efficiency.

In many AI-based numerical weather prediction and climate model emulation studies, it has been shown to improve stability and overall emulator performance to perform a short rollout of each batch during training, and evaluate the loss function over this short rollout. For FloeNet, we find that rolling out for 8 timesteps (2 days) leads to optimal performance, although, for computational reasons, we did not test more than 8. For the loss function itself, we use the standard mean-squared error (MSE) metric, averaged across all output channels. It is important to note that we apply an increased weighting to \texttt{siconc}, \texttt{simass}, and \texttt{sisnmass} in the loss function (see Table \ref{tab:tab1}), which we find is critical for stable training when using our mass budget approach. Our final model then corresponds to the set of weights (checkpoint) for which the channel-mean validation loss is lowest across the 50-epoch training period. We refer the reader to Table \ref{tab:tab2} for other specific hyperparameters related to training FloeNet.

\section{Sensitivity tests}

\subsection*{Random seed}
\begin{figure}[t!]
    \centering
    \includegraphics[width=1\linewidth]{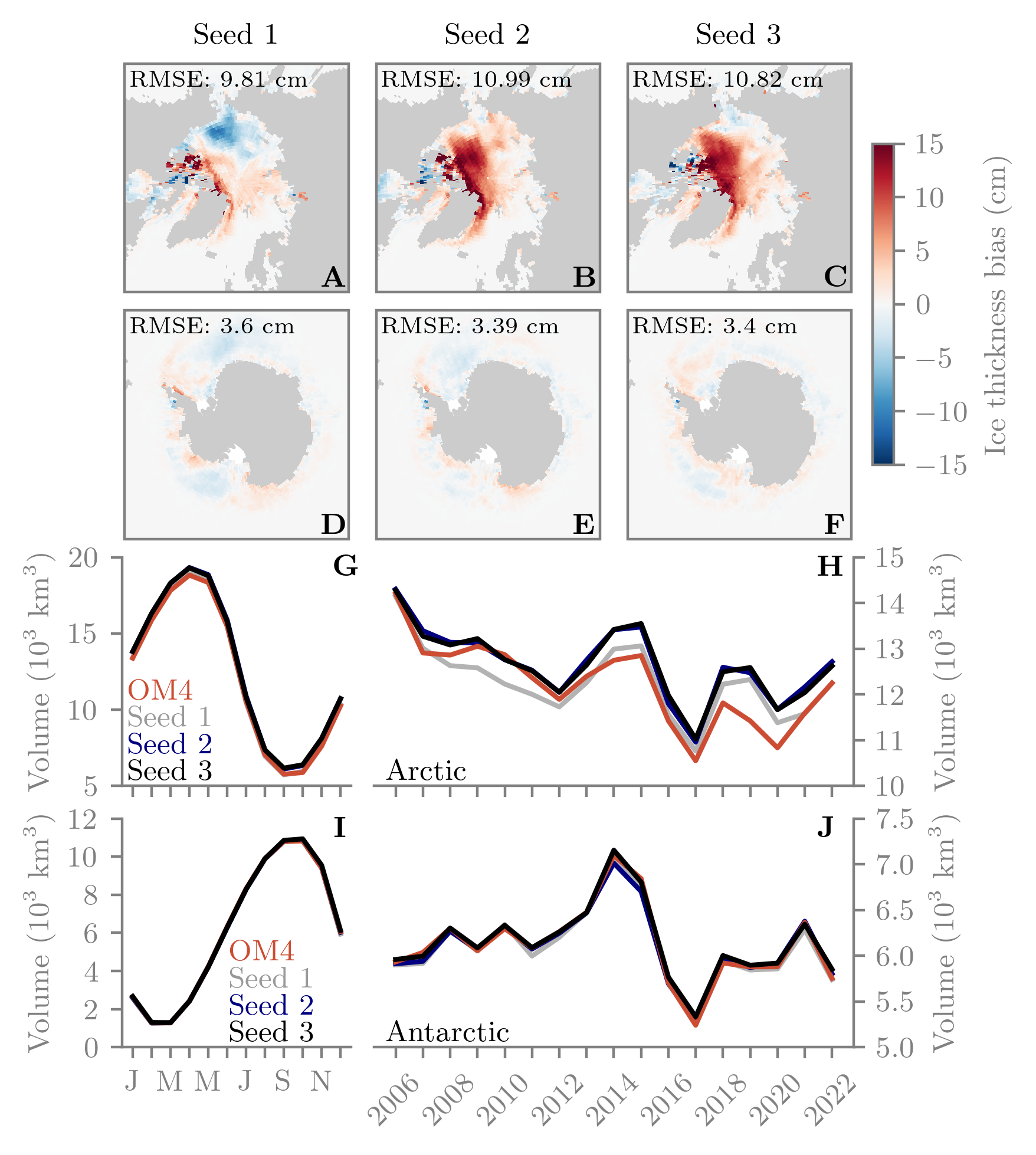}
    \caption{Validation of sea ice thickness and volume for different random seed initializations during training. (\textbf{A}--\textbf{C}) Arctic sea ice thickness bias. (\textbf{D}--\textbf{F}) Same as (\textbf{A}--\textbf{C}) but for Antarctic. (\textbf{G}) Pan-Arctic sea ice volume. (\textbf{H}) Annual-mean Arctic sea ice volume between 2006--2022. (\textbf{I},\textbf{J}) Same as (\textbf{G},\textbf{H}) but for Antarctic.}
    \label{fig:randomseed_ice}
\end{figure}

\begin{figure}[t!]
    \centering
    \includegraphics[width=1\linewidth]{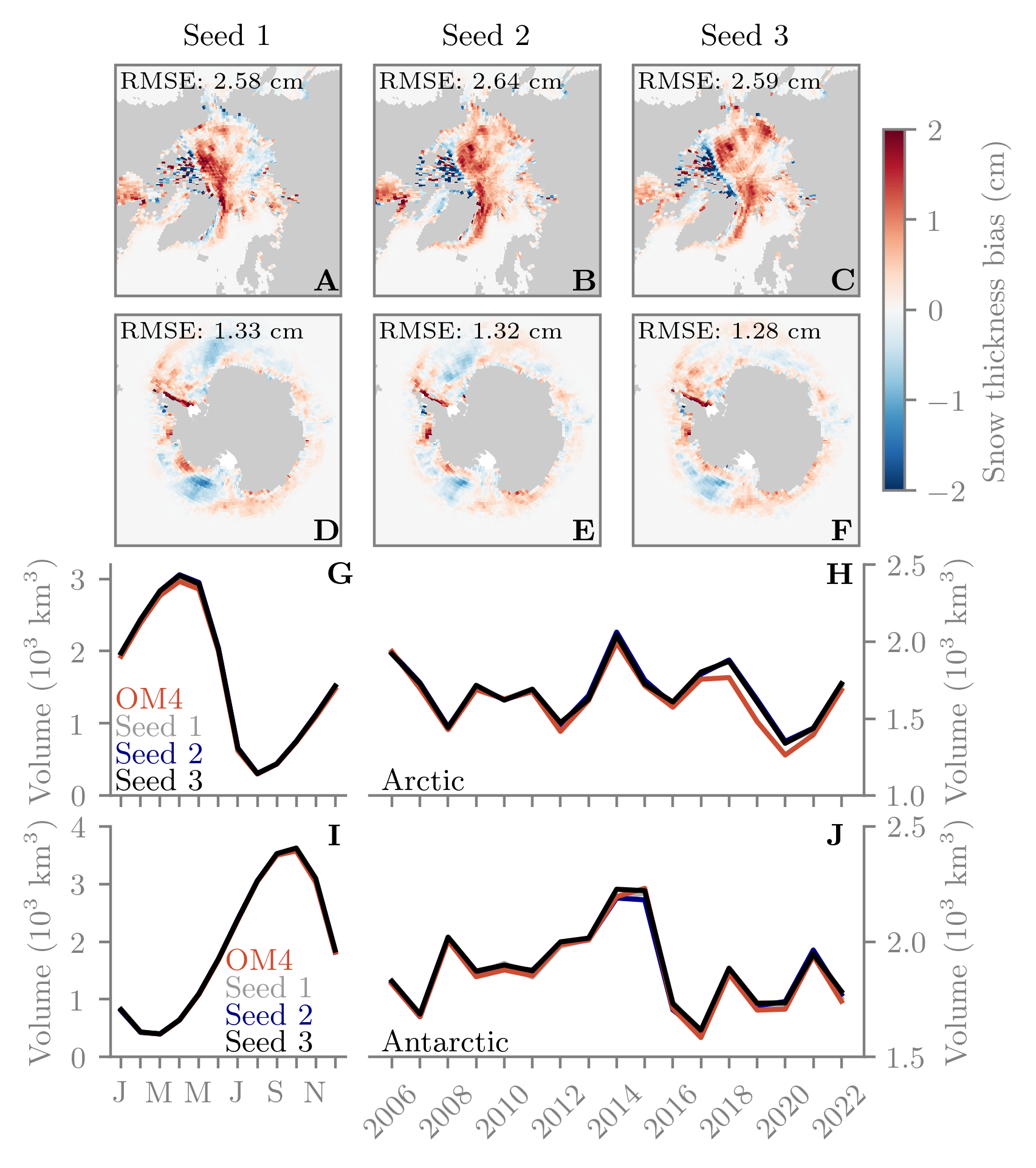}
    \caption{Same as Fig. S9 but for snow-on-sea-ice thickness and volume.}
    \label{fig:randomseed_snow}
\end{figure}

In Fig. \ref{fig:randomseed_ice} we show the average sea ice thickness bias relative to OM4 over the period 2006--2022, for three different random seed initializations of the network weights during training (Figs. \ref{fig:randomseed_ice}A--\ref{fig:randomseed_ice}F). We also show the seasonal cycle climatology over the same period for the Arctic (Fig. \ref{fig:randomseed_ice}G) and Antarctic (Fig. S9I), and the annual-mean sea ice volume time series (Fig. \ref{fig:randomseed_ice}H and \ref{fig:randomseed_ice}J). From this figure we can see that FloeNet's sea ice thickness and volume are somewhat sensitive to random seed initialization, and more so in the Arctic than Antarctic. The RMSE ranges from 9.81 cm to 10.99 cm in the Arctic and 3.39 cm to 3.60 cm in the Antarctic. For snow-on-sea-ice (Fig. \ref{fig:randomseed_snow}), there is less sensitivity to random seed, with RMSE ranging from 2.58 cm to 2.64 cm in the Arctic and 1.28 cm to 1.33 cm in the Antarctic. We note that running many more random seed tests would help to constrain the uncertainty on a given model's predictions. However, due to the fact that our model takes $\sim$5 days to train, we stopped at 3 random seeds. For each of the sensitivity tests below, we therefore only train once, initializing weights using the same random seed across all tests.

\subsection*{Processor mesh resolution}

\begin{figure}[t!]
    \centering
    \includegraphics[width=1\linewidth]{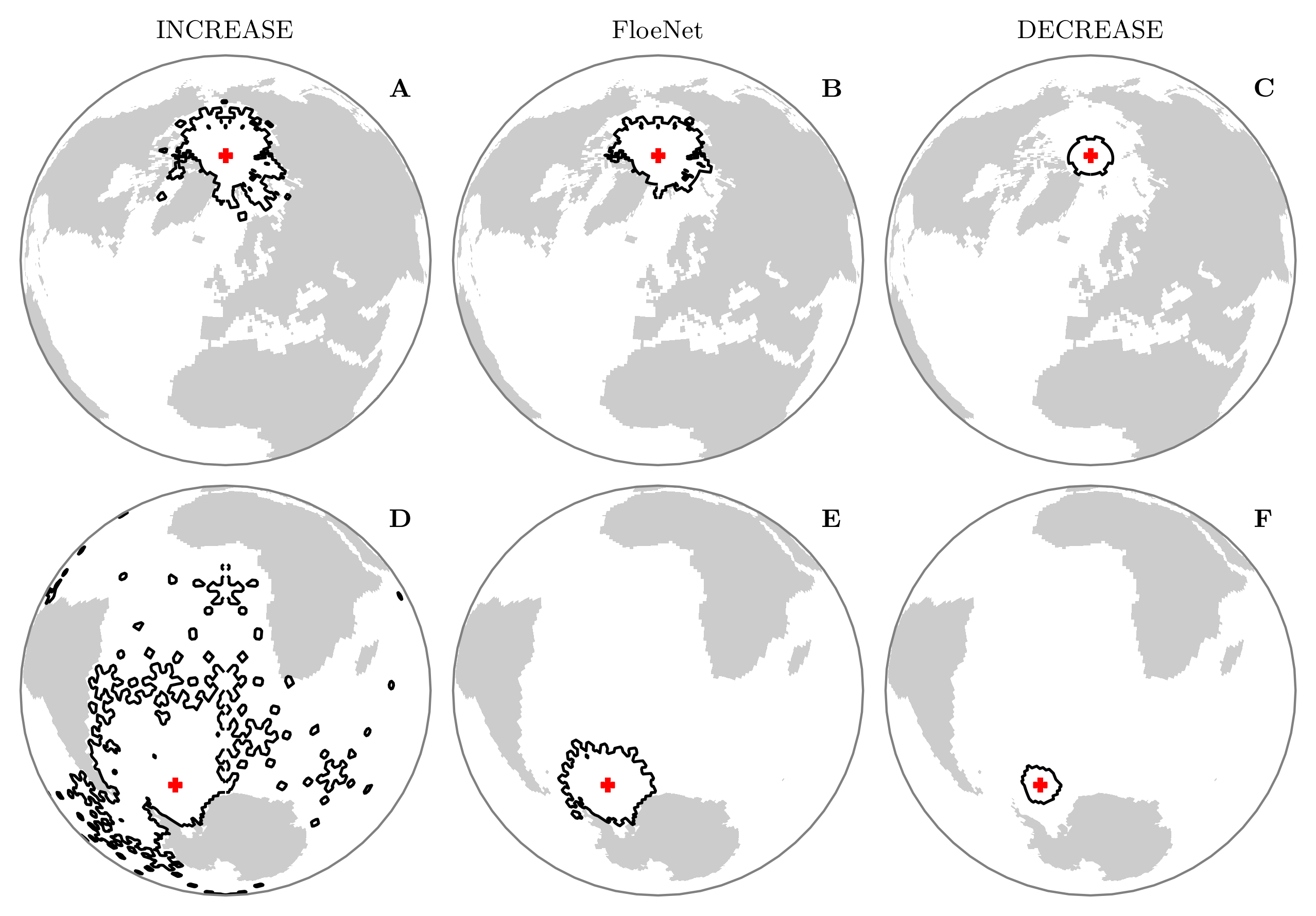}
    \caption{Receptive field of different multi-mesh resolutions, shown as contours of grid cells which are activated from a given seed node (red) in the Arctic (\textbf{A}--\textbf{C}) and Antarctic (\textbf{D}--\textbf{F}).  (\textbf{A},\textbf{D}) All resolutions M$_0$ through M$_6$. (\textbf{B},\textbf{E}) Resolutions M$_4$ through M$_6$ (FloeNet). (\textbf{C},\textbf{F}) Only M$_6$. For scale, the contours in (\textbf{B}) and (\textbf{E}) have a radius of approximately 1500 km, while those in (\textbf{C}) and (\textbf{F}) have a radius of approximately 700 km.}
    \label{fig:receptivefield}
\end{figure}

In this section, we show examples of increasing and decreasing FloeNet's receptive field by changing the resolution of the processor multi-mesh. Specifically, we test an `increasing' receptive field case by including all mesh resolutions from M$_0$ through to M$_6$, as well as `decreasing' case which only uses M$_6$ (see Fig. \ref{fig:receptivefield}). Due to the fact that mesh nodes are filtered over land, the physical size of the receptive field for the increasing case varies between the Arctic and Antarctic. For example, a seed node in the Southern Ocean has a much larger receptive field because it can connect to coarser-resolution mesh nodes over the ocean (Fig. \ref{fig:receptivefield}D). Meanwhile in the Arctic, some of the coarser-resolution mesh nodes (including M$_0$) fall over land, so a seed node in the center of the Arctic Ocean cannot propagate information as far (Fig. \ref{fig:receptivefield}A). The receptive field for FloeNet's default configuration is approximately 1500 km (Fig. \ref{fig:receptivefield}B and \ref{fig:receptivefield}E), and in the decreasing case is 700 km (Fig. \ref{fig:receptivefield}C and \ref{fig:receptivefield}F).

Varying the mesh resolution provides a cost-effective way to test for receptive field sensitivity, as changing the mesh resolution does not change the number of weights of the network. We note that changing the number of stacked GNN layers in the processor can also change the receptive field, however this can increase the number of weights and computational cost considerably. We therefore use 4 stacked GNN layers in each test here.

\begin{figure}[t!]
    \centering
    \includegraphics[width=1\linewidth]{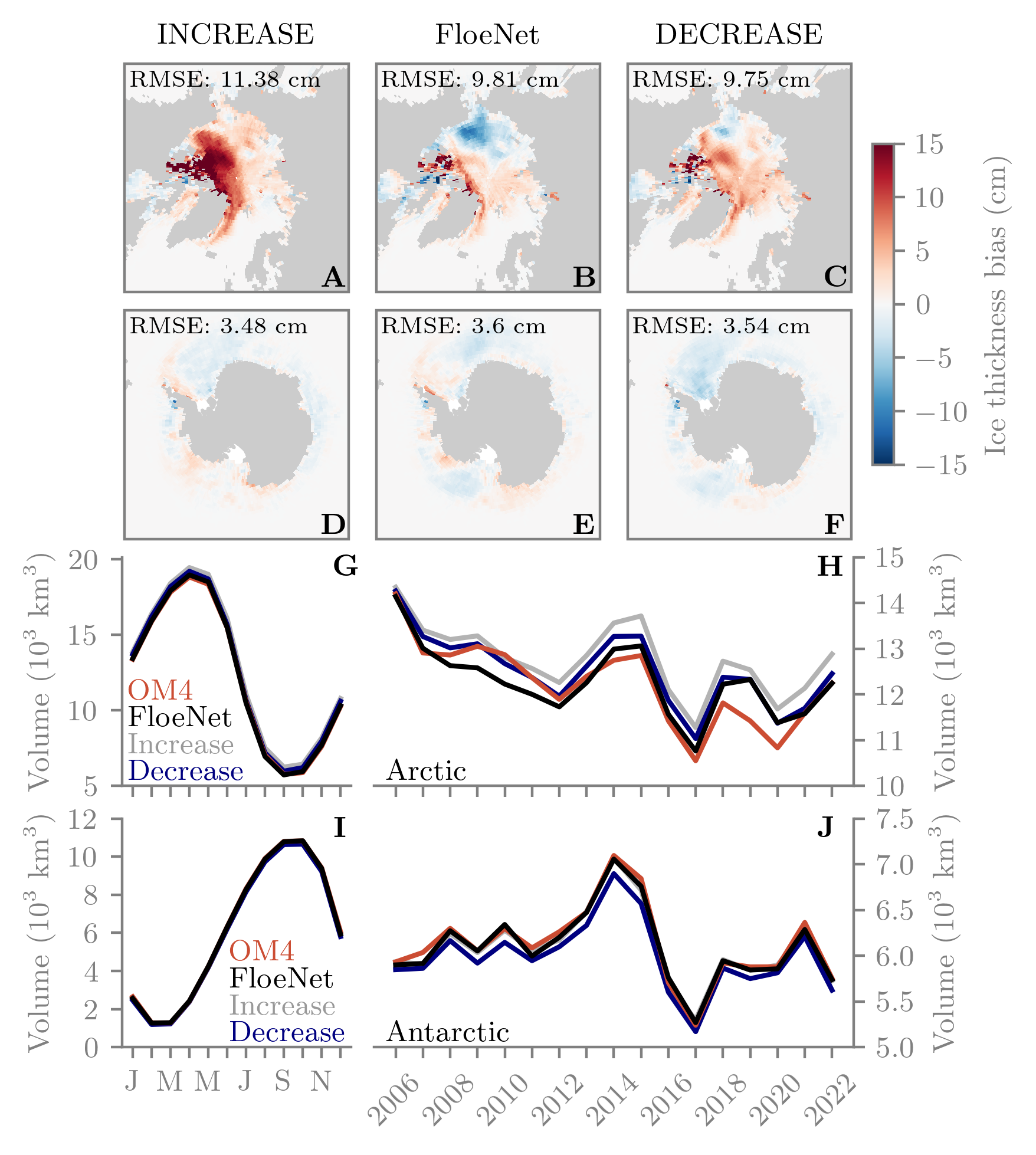}
    \caption{Validation of sea ice thickness and volume for different receptive fields; see Fig. S11 for definitions. (\textbf{A}--\textbf{C}) Arctic sea ice thickness bias for increased mesh receptive field, the default FloeNet mesh, and a smaller receptive field, respectively. (\textbf{D}--\textbf{F}) Same as (\textbf{A}--\textbf{C}) but for Antarctic. (\textbf{G}) Pan-Arctic sea ice volume. (\textbf{H}) Annual-mean Arctic sea ice volume between 2006--2022. (\textbf{I},\textbf{J}) Same as (\textbf{G},\textbf{H}) but for Antarctic.}
    \label{fig:receptivefield_rollouts_ice}
\end{figure}

\begin{figure}[t!]
    \centering
    \includegraphics[width=1\linewidth]{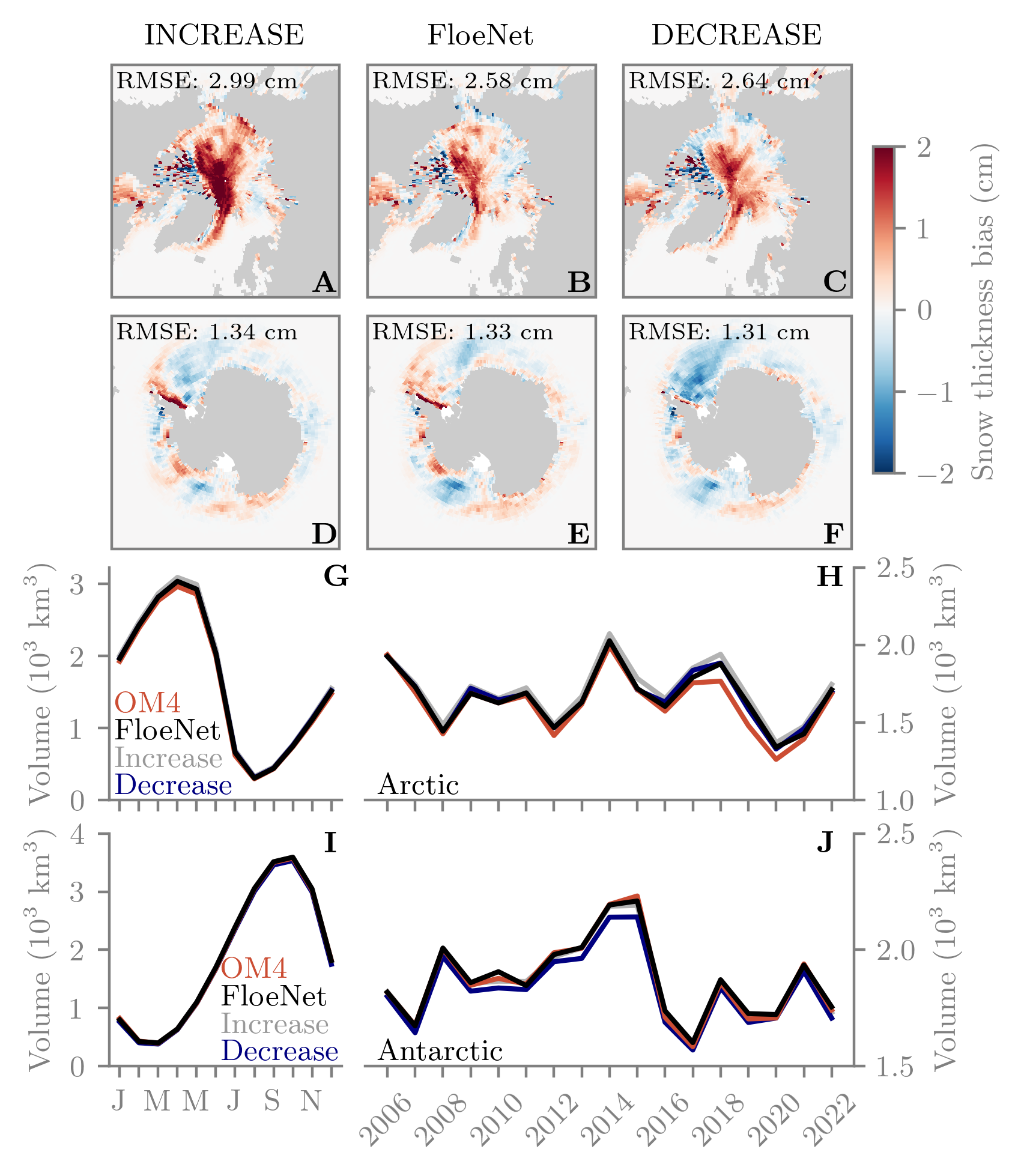}
    \caption{Same as Fig. S13 but for snow-on-sea-ice thickness and volume.}
    \label{fig:receptivefield_rollouts_snow}
\end{figure}

In Fig. \ref{fig:receptivefield_rollouts_ice} we show the average sea ice thickness bias relative to OM4 over the period 2006--2022, for the increasing case (Figs. \ref{fig:receptivefield_rollouts_ice}A, \ref{fig:receptivefield_rollouts_ice}D), default FloeNet (Figs. \ref{fig:receptivefield_rollouts_ice}B, \ref{fig:receptivefield_rollouts_ice}E), and decreasing case (Figs. \ref{fig:receptivefield_rollouts_ice}C, \ref{fig:receptivefield_rollouts_ice}F). We also show the seasonal cycle climatology over the same period for the Arctic (Fig. \ref{fig:receptivefield_rollouts_ice}G) and Antarctic (Fig \ref{fig:receptivefield_rollouts_ice}I), and the annual-mean sea ice volume time series (Figs. \ref{fig:receptivefield_rollouts_ice}H and \ref{fig:receptivefield_rollouts_ice}J). These results indicate that a smaller receptive field generally leads to improved generalization. However, it is important to note that sea ice thickness and volume predictions are sensitive to random seed initialization of the network weights during training (see above). Therefore, these results (especially the default FloeNet vs decreasing case) may not be significantly different. See Fig. \ref{fig:receptivefield_rollouts_snow} for similar results, but for snow-on-sea-ice.

\subsection*{Ablation of forcing variables}

Due to computational restrictions, we do not provide an exhaustive set of ablation experiments for FloeNet. However, our evaluation of the full-state model in the main text highlights the importance of emulating mass and area budget terms. In this section, we provide two ablation experiments, where we remove input forcing variables: rate of snow fall (\texttt{SNOWFL}) and ice-ocean stress terms (\texttt{tauuo}, \texttt{tauvo}). Our motivation for choosing these variables is that \texttt{SNOWFL} is currently not predicted by the ACE2 atmospheric emulator. Similarly, \texttt{tauuo} and \texttt{tauvo} are currently not predicted by the Samudra ocean emulator. Therefore, if FloeNet were to be coupled to either of these components, they would need to be re-trained to output these terms (\texttt{BHEAT} is also not directly output from Samudra, but this can be directly diagnosed using the predicted surface temperature and salinity fields).

\begin{figure}[t!]
    \centering
    \includegraphics[width=\linewidth]{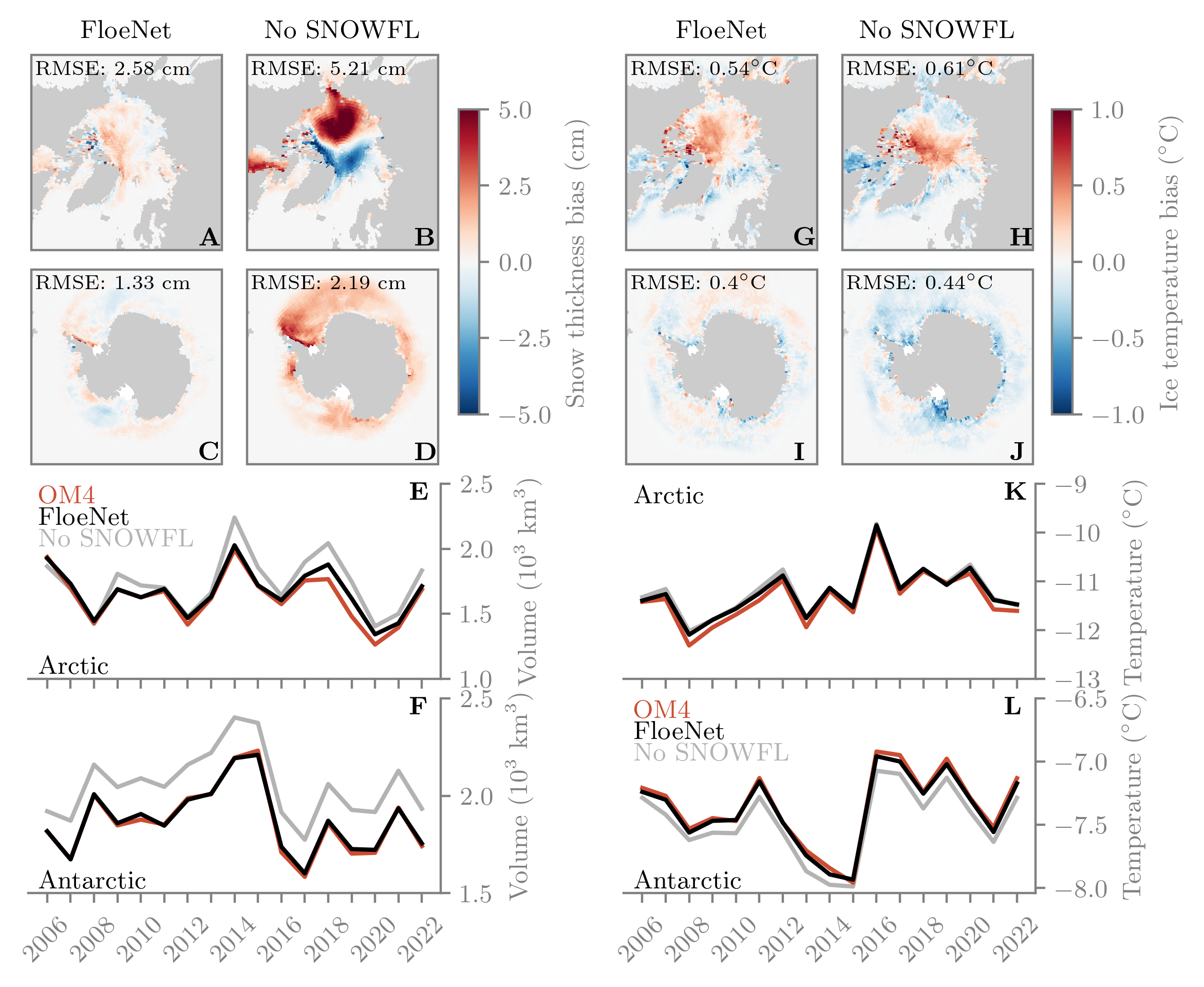}
    \caption{Ablation test, removing rate of snowfall (\texttt{SNOWFL}) from the set of input forcing variables. (\textbf{A}--\textbf{D}) Arctic and Antarctic snow-on-sea-ice thickness biases. (\textbf{E},\textbf{F}) Annual-mean snow volume time series between 2006--2022. (\textbf{G}--\textbf{L}) Same as (\textbf{A}--\textbf{F}) but for ice-surface skin temperature.}
    \label{fig:noSNOWFL}
\end{figure}

Fig. \ref{fig:noSNOWFL} shows, as expected, removing \texttt{SNOWFL} leads to large biases in snow thickness, with an approximately 2$\times$ increase in RMSE in the Arctic and Antarctic. The bias does not manifest as severely in the Arctic annual-mean snow volume time series, which is due to compensating positive and negative biases in the Pacific and Atlantic sectors, respectively. We also see a slight increase in the ice-surface skin temperature bias when removing \texttt{SNOWFL}, and the spatial pattern of these temperature bias changes is consistent with the expected impact of the snow biases. However, more random seed initializations would be needed to determine the significance of this bias change.  

\begin{figure}[t!]
    \centering
    \includegraphics[width=\linewidth]{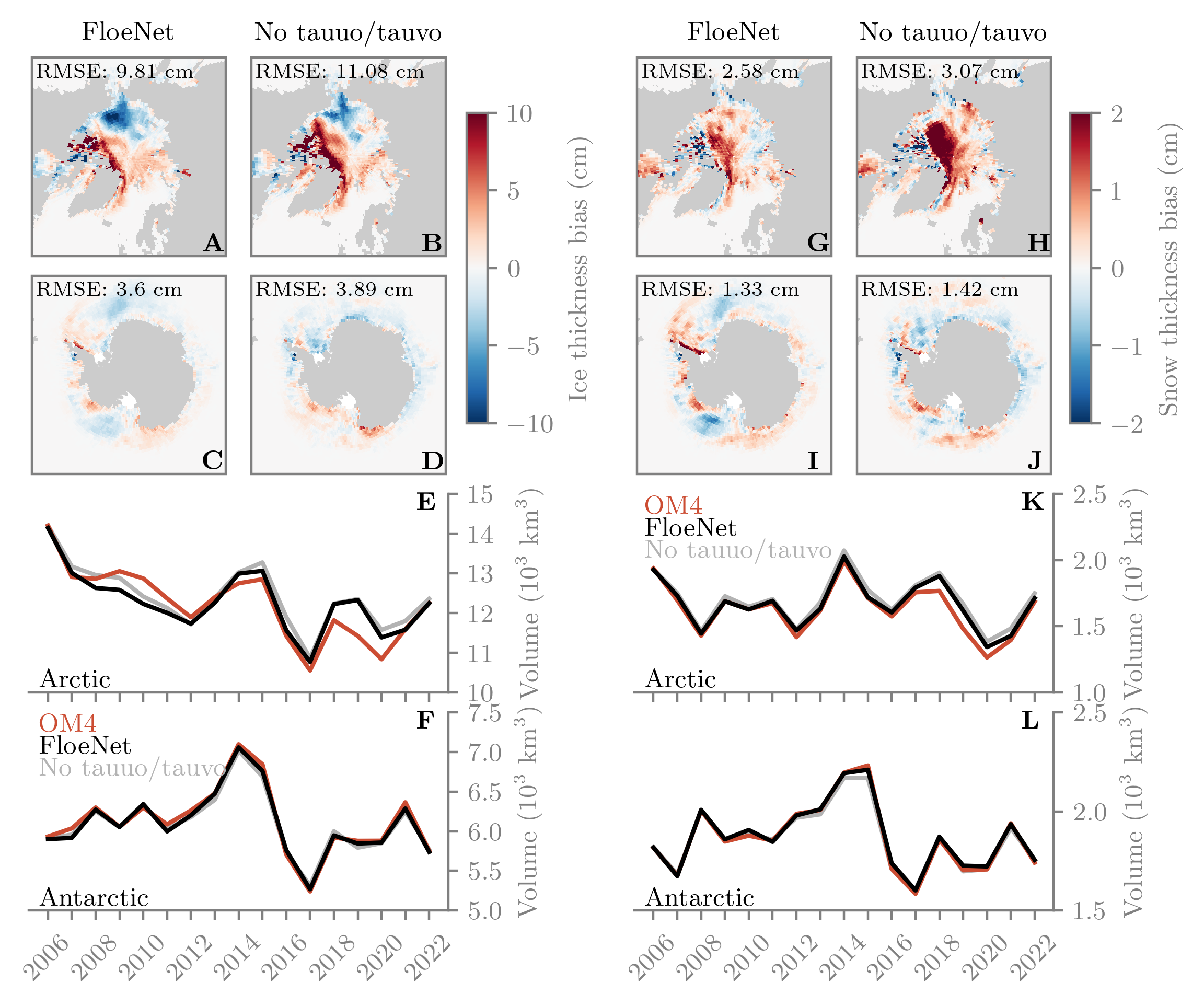}
    \caption{Ablation test, removing ice-ocean stress (\texttt{tauuo},\texttt{tauvo}) from the set of input forcing variables. (\textbf{A}--\textbf{D}) Arctic and Antarctic sea ice thickness biases. (\textbf{E},\textbf{F}) Annual-mean sea ice volume time series between 2006--2022. (\textbf{G}--\textbf{L}) Same as (\textbf{A}--\textbf{F}) but for snow-on-sea-ice thickness and volume.}
    \label{fig:notau}
\end{figure}

Removing ice-ocean stress terms (Fig. \ref{fig:notau}) has subtle impacts on sea ice and snow-on-sea-ice thickness, with increases in RMSE typically $\leq$ 1 cm in both cases. Some interesting features include an increase in sea ice thickness bias along the area north of Greenland and the Canadian Arctic Archipelago (Fig. \ref{fig:notau}B). This is where sea ice is thickest and less mobile, as it `piles up' along the coastal margin. We would therefore expect ice-ocean stress to be informative of properties like sea ice mass transport, as this encodes a negative feedback that arises due to increased ice-ocean stress under increased sea ice drift speeds. We see a similar bias increase in snow thickness in this region (Fig. \ref{fig:notau}H), although the bias extends further north into the central Arctic Ocean. This may be an indirect effect of the degradations in ice thickness.

\subsection*{6-hour vs 5-day timestep}

\begin{figure}[t!]
    \centering
    \includegraphics[width=\linewidth]{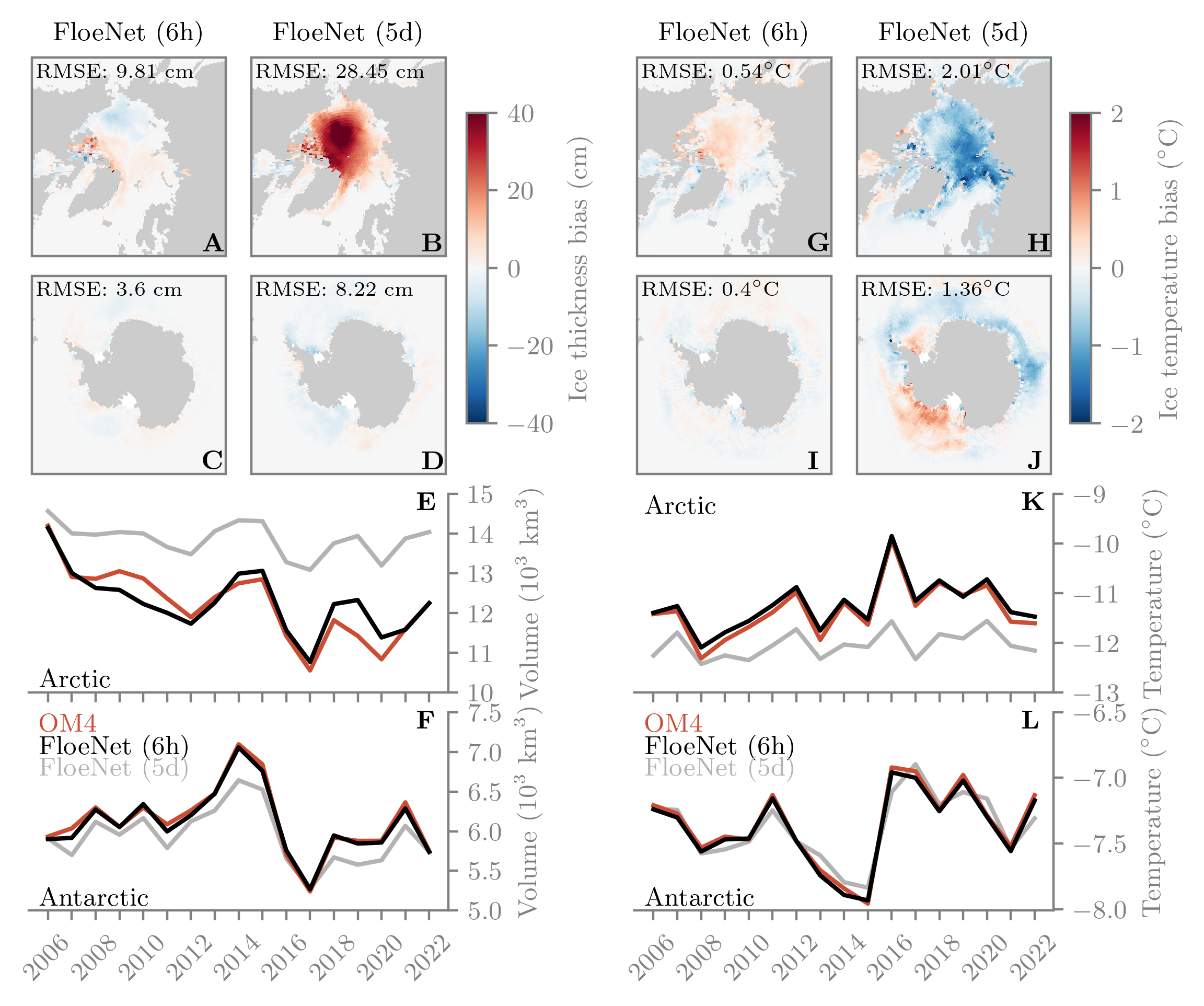}
    \caption{Evaluating FloeNet rollouts when trained with a 6-hour vs 5-day timestep. (\textbf{A}--\textbf{D}) Arctic and Antarctic sea ice thickness biases. (\textbf{E},\textbf{F}) Annual-mean sea ice volume time series between 2006--2022. (\textbf{G}--\textbf{L}) Same as (\textbf{A}--\textbf{F}) but for ice-surface skin temperature.}
    \label{fig:6hv5d}
\end{figure}

In Fig. \ref{fig:6hv5d} we show the effect of coarsening FloeNet's 6-hour timestep to 5 days. We choose 5 days as this is the current timestep of the ocean component of the SamudrACE coupled emulator, which includes sea ice. In Figs. \ref{fig:6hv5d}A--\ref{fig:6hv5d}F we show that a 5-day timestep significantly degrades sea ice thickness and volume predictions, particularly in the Arctic. The 5-day timestep model has a large positive thickness bias across the Arctic Ocean, with an 18.64 cm increase in RMSE. In the sea ice volume time series (Fig. \ref{fig:6hv5d}E), we can see that the 5-day model has lower inter-annual variability and does not reproduce the downward trend in volume over the 2006--2022 period. For ice-surface skin temperature (Figs. \ref{fig:6hv5d}G--\ref{fig:6hv5d}L), the 5-day model does a reasonable job at reproducing the Antarctic mean state and variability (Fig. \ref{fig:6hv5d}L), although shows a systematic low bias and lower variability in the Arctic (Fig. \ref{fig:6hv5d}K).

\subsection*{Architecture}
In our final sensitivity test, we compare our FloeNet GNN architecture to a model which uses a ConvNeXt architecture. The ConvNeXt model is a convolution-based UNet architecture, which forms the basis of the Samudra ocean emulator (Dheeshjith et al., 2025) and SamudraI ocean and sea ice emulator (Duncan et al., 2025). We use a smaller version of the model compared to Samudra and SamudraI, with  approximately the same number of weights as FloeNet ($\sim$3 million); we did test a larger ConvNeXt model with 13.5 million weights, however this performs worse than the model with $\sim$3 million weights (not shown). For our small ConvNeXt model, we use a 3-layer network, with channel widths equal to 64, 128, and 256 in each layer. The dilation rates of each layer are then 1, 2, and 4. The inputs and outputs, hyperparameters, and mass budget conservation approach of the ConvNeXt model are identical to FloeNet. The ConvNeXt model is also trained with a 6-hour timestep and for 50 epochs.

\begin{figure}[t!]
    \centering
    \includegraphics[width=\linewidth]{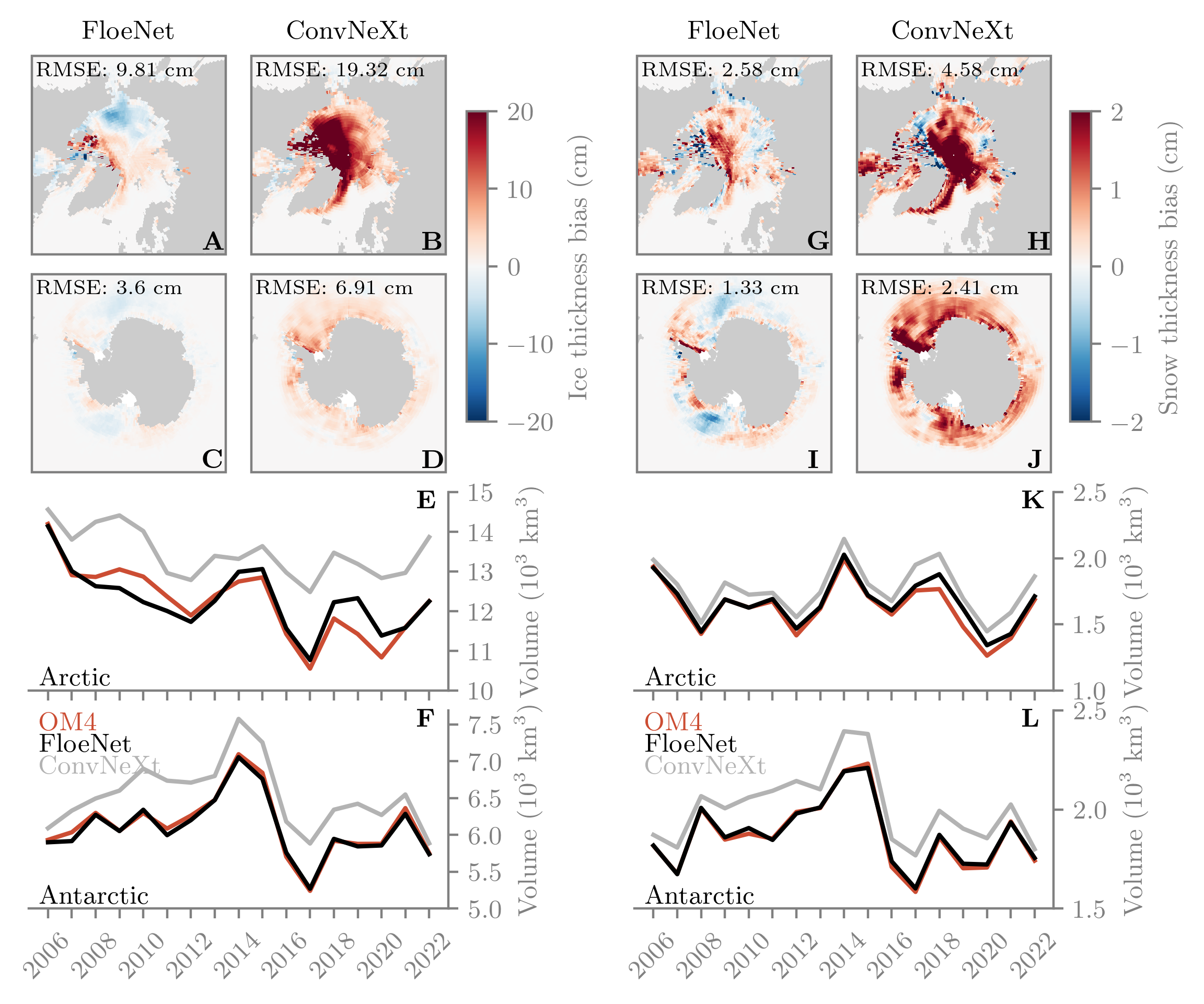}
    \caption{Comparing Graph- (FloeNet) and Convolution-based (ConvNeXt) architectures. (\textbf{A}--\textbf{D}) Arctic and Antarctic sea ice thickness biases. (\textbf{E},\textbf{F}) Annual-mean sea ice volume time series between 2006--2022. (\textbf{G}--\textbf{L}) Same as (\textbf{A}--\textbf{F}) but for snow-on-sea-ice thickness and volume.}
    \label{fig:convnext}
\end{figure}

Fig. \ref{fig:convnext} shows the mean sea ice thickness and snow-on-sea-ice thickness biases, as well as the annual-mean volume time series of each quantity. We notice here that the ice and snow thickness is significantly degraded using the ConvNeXt model, with a 9.51 cm and 2 cm increase in Arctic sea ice and snow thickness RMSE, respectively. In the Antarctic, ice and snow thickness RMSE also increases by 3.31 cm and 1.08 cm, respectively. In the Arctic, the ConvNeXt model shows lower annual-mean sea ice volume variability and underestimation of the 17-year trend. This aligns with out-of-sample testing of the Samudra ocean emulator, which tends to show lower ocean temperature variability and underestimation of warming trends (Dheeshjith et al., 2025).

\begin{table}[h!]
\center
\caption{Details of the FloeNet input and output variables. Radiative fluxes are defined at the top of the sea ice. }
\begin{tabular}{p{1.5cm} p{6cm} p{1.75cm} p{2.5cm} p{2.25cm}} \textbf{Name} & \textbf{Description} & \textbf{Type} & \textbf{Time def.} & \textbf{Loss weight} \\
\hline
\texttt{siconc} & sea ice area fraction [0-1] & prognostic & 6-hour snapshot & 5 \\
\texttt{simass} & sea ice mass [kgm$^{-2}$] & prognostic & 6-hour snapshot & 10 \\
\texttt{sisnmass} & snow mass [kgm$^{-2}$] & prognostic & 6-hour snapshot & 5 \\
\texttt{TS} & ice-surface skin temperature [$^\circ$C] & prognostic & 6-hour snapshot & 1 \\
\texttt{LSRCi} & sea ice mass source [kgm$^{-2}$s$^{-1}$] & prognostic & 6-hour mean & 1\\
\texttt{LSNKi} & sea ice mass sink [kgm$^{-2}$s$^{-1}$] & prognostic & 6-hour mean & 1\\
\texttt{XPRTi} & sea ice mass transport [kgm$^{-2}$s$^{-1}$] & prognostic & 6-hour mean & 1 \\
\texttt{LSRCs} & snow mass source [kgm$^{-2}$s$^{-1}$] & prognostic & 6-hour mean & 1 \\
\texttt{LSNKs} & snow mass sink [kgm$^{-2}$s$^{-1}$] & prognostic & 6-hour mean & 1 \\
\texttt{XPRTs} & snow mass transport [kgm$^{-2}$s$^{-1}$] & prognostic & 6-hour mean & 1 \\
\texttt{LSRCc} & sea ice area source [s$^{-1}$] & prognostic & 6-hour mean & 1 \\
\texttt{LSNKc} & sea ice area sink [s$^{-1}$] & prognostic & 6-hour mean  & 1\\
\texttt{XPRTc} & sea ice area transport [s$^{-1}$] & prognostic & 6-hour mean & 1 \\
\texttt{SALTF} & ice-to-ocean salt flux [kgm$^{-2}$s$^{-1}$] & prognostic & 6-hour mean & 1 \\
\texttt{TMELT} & top melting energy flux [Wm$^{-2}$] & diagnostic & 6-hour mean & 1 \\
\texttt{BMELT} & bottom melting energy flux [Wm$^{-2}$] & diagnostic & 6-hour mean & 1 \\
\texttt{SWDN} & downward shortwave [Wm$^{-2}$] & forcing & 6-hour mean & -- \\
\texttt{LWDN} & downward longwave [Wm$^{-2}$] & forcing & 6-hour mean & -- \\
\texttt{SH} & sensible heat flux [Wm$^{-2}$] & forcing & 6-hour mean & -- \\
\texttt{LH} & latent heat flux [Wm$^{-2}$] & forcing & 6-hour mean & -- \\
\texttt{SNOWFL} & rate of snow fall [kgm$^{-2}$s$^{-1}$] & forcing & 6-hour mean & -- \\
\texttt{FA\_X} & zonal wind stress [Pa] & forcing & 6-hour mean & -- \\
\texttt{FA\_Y} & meridional wind stress [Pa] & forcing & 6-hour mean & -- \\
\texttt{BHEAT} & ocean-to-ice heat flux [Wm$^{-2}$] & forcing & 5-day mean & -- \\
\texttt{SSH} & sea-surface height [m] & forcing & 5-day mean & -- \\
\texttt{tauuo} & zonal ice-ocean stress [Pa] & forcing & 5-day mean & -- \\
\texttt{tauvo} & meridional ice-ocean stress [Pa] & forcing & 5-day mean & -- \\
\end{tabular}
\label{tab:tab1}
\end{table}

\vspace{40pt}

\begin{table}[h!]
\center
\caption{Details of the FloeNet model and training configuration.}
\begin{tabular}{p{5cm} | p{4cm}} \\
\hline
Activation functions & SiLU \\
Batch size & 32 \\
Bias parameters & True \\
EMA decay & 0.999 \\
Epochs & 50 \\
Latent/feature space dimension & 256 \\
Learning rate & 0.0001 \\
Loss function & Mean-squared error \\
Normalization (inputs and outputs) & Zero-mean and unit variance \\
Normalization (activations) & Layer-wise \\
Number of model parameters & 3,310,608 \\
Optimizer & AdamW \\
Weight decay & 0.01 \\
\end{tabular}
\label{tab:tab2}
\end{table}

\begin{figure}[t!]
    \centering
    \includegraphics[width=1\linewidth]{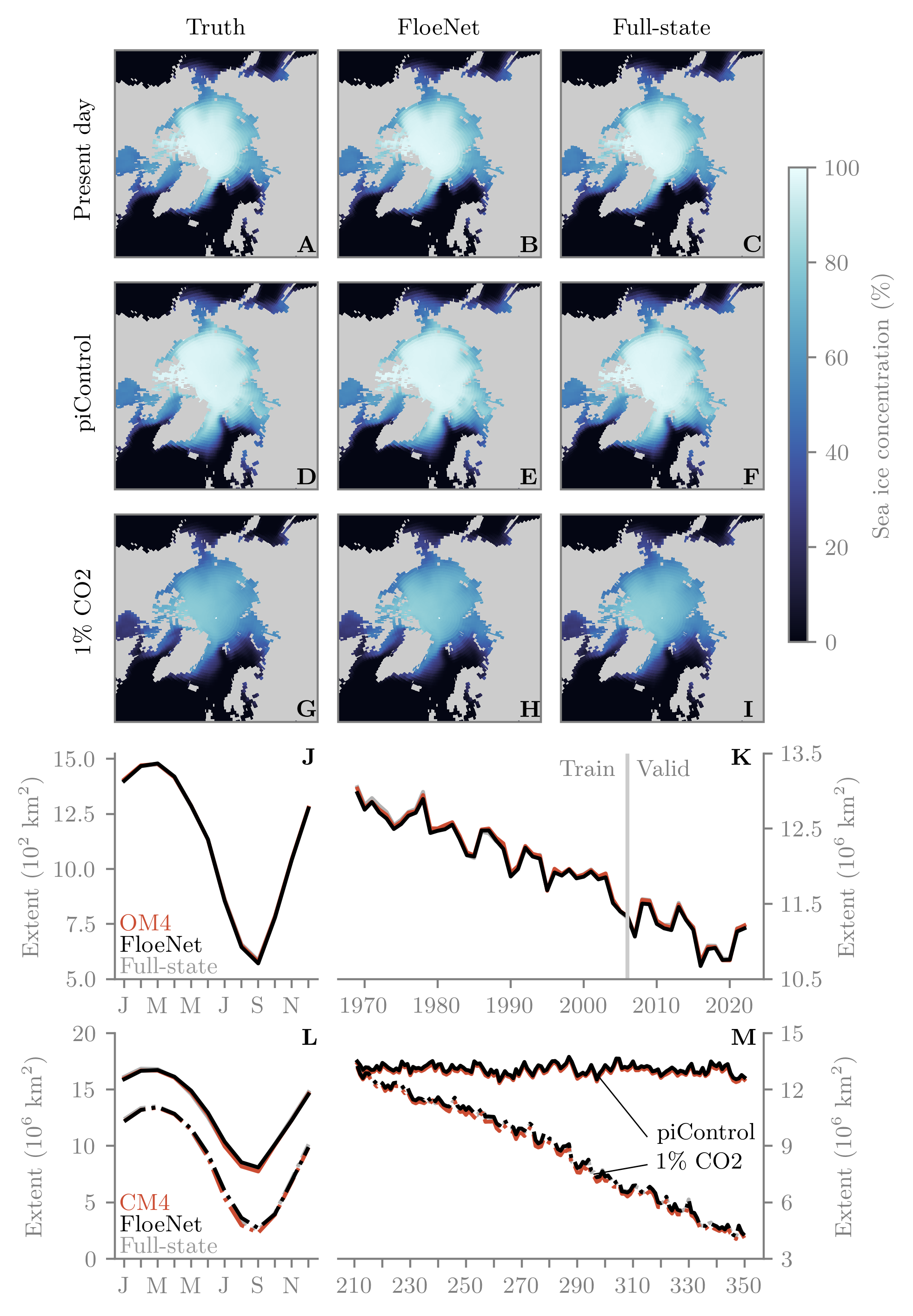}
    \caption{Same as Fig. 2 of main article, but for Arctic sea ice concentration and extent.}
    \label{fig:ArcticExtent}
\end{figure}

\begin{figure}[t!]
    \centering
    \includegraphics[width=1\linewidth]{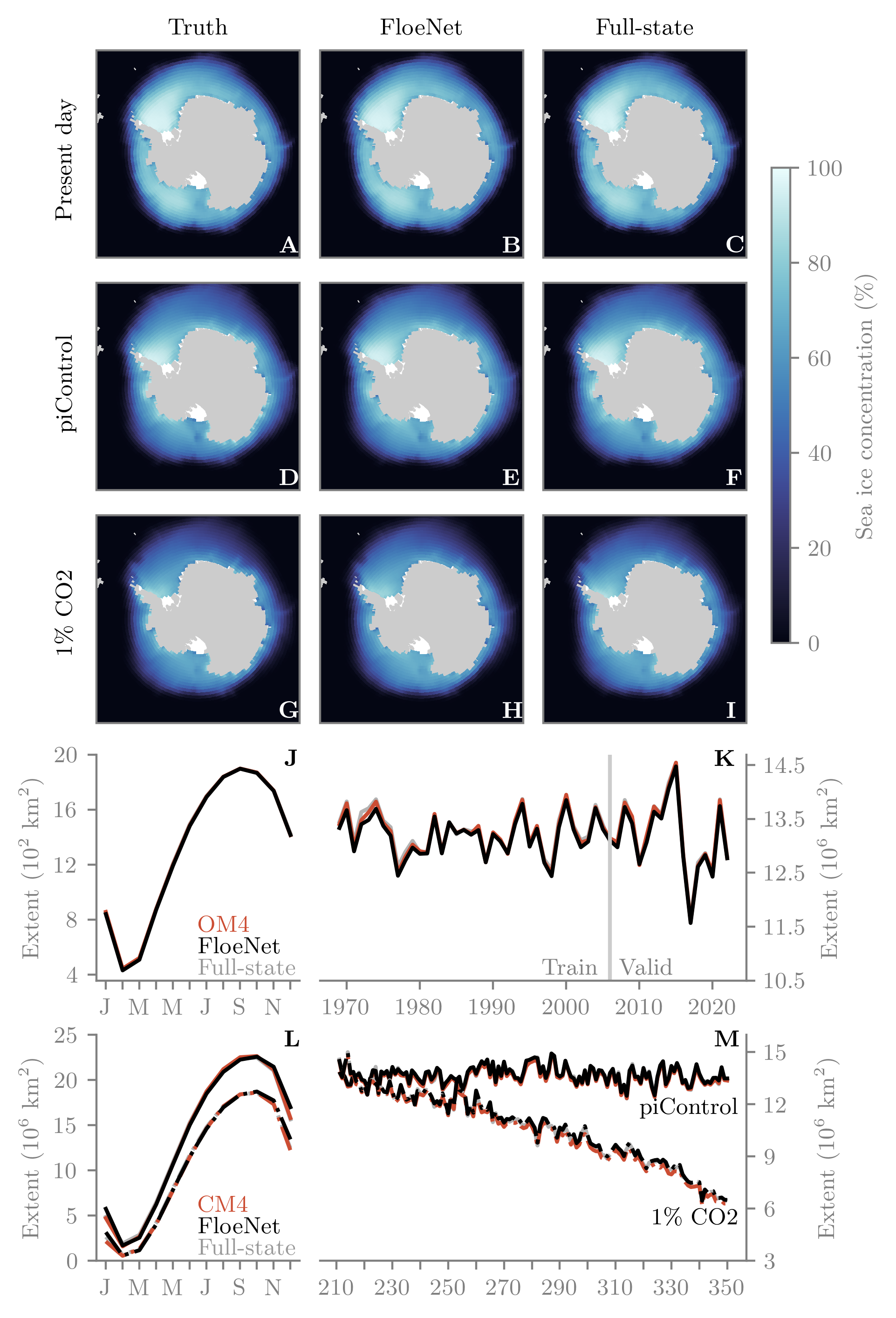}
    \caption{Same as Fig. 2 of main article, but for Antarctic sea ice concentration and extent.}
    \label{fig:AntarcticExtent}
\end{figure}

\begin{figure}[t!]
    \centering
    \includegraphics[width=\linewidth]{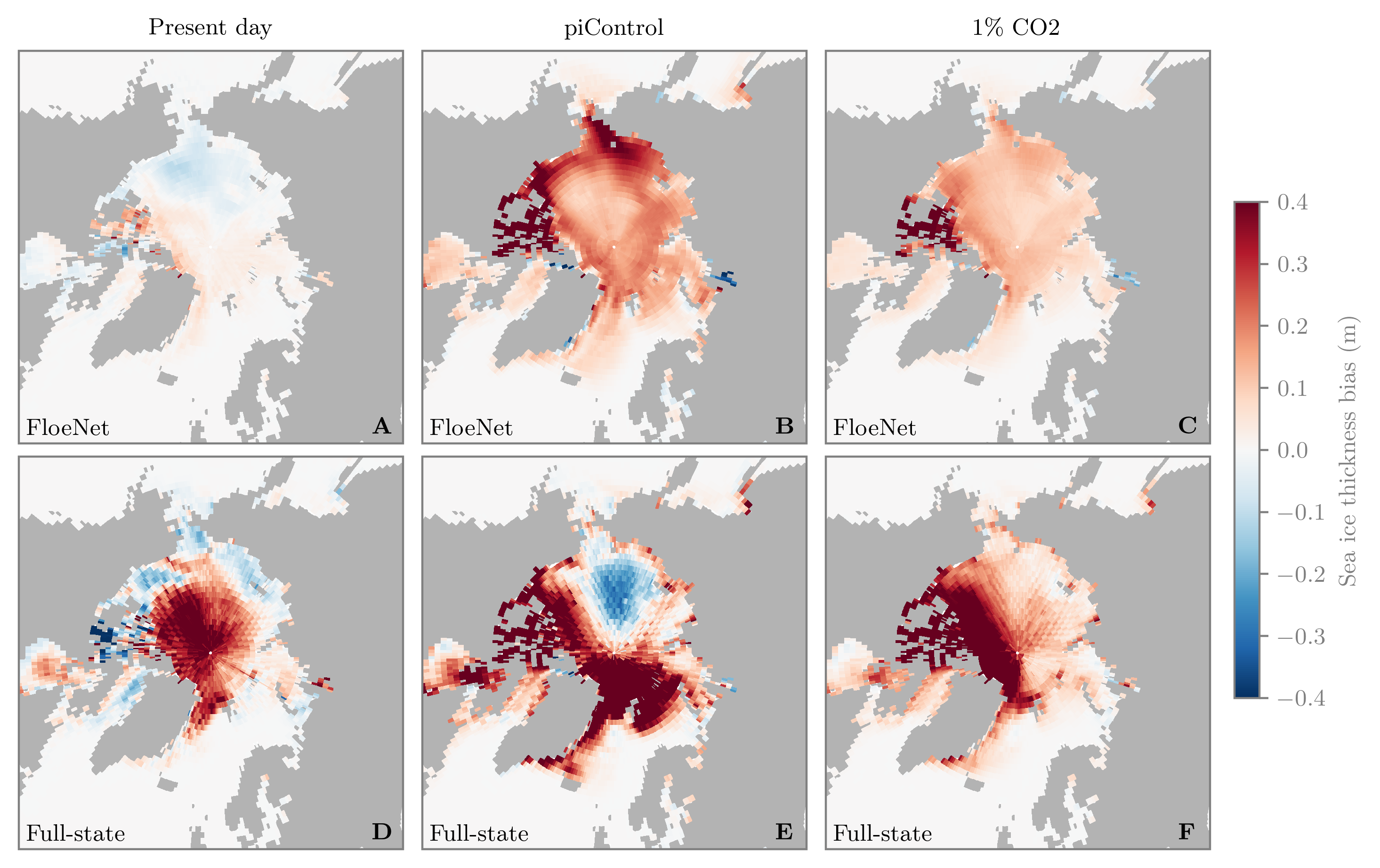}
    \caption{Time-mean Arctic sea ice thickness bias under different forcings. (\textbf{A}) FloeNet's Arctic sea ice thickness bias under present-day (OM4) forcing, for the period 2006--2022. (\textbf{B}) Same as (\textbf{A}) but under piControl forcing, for the period 211--350. (\textbf{C}) Same as (\textbf{B}) but under 1\% CO\textsubscript{2} forcing, for the period 211--350. (\textbf{D}--\textbf{F}) Same as (\textbf{A}--\textbf{C}) but for the full-state model.}
    \label{fig:ArcticBias}
\end{figure}

\begin{figure}[t!]
    \centering
    \includegraphics[width=\linewidth]{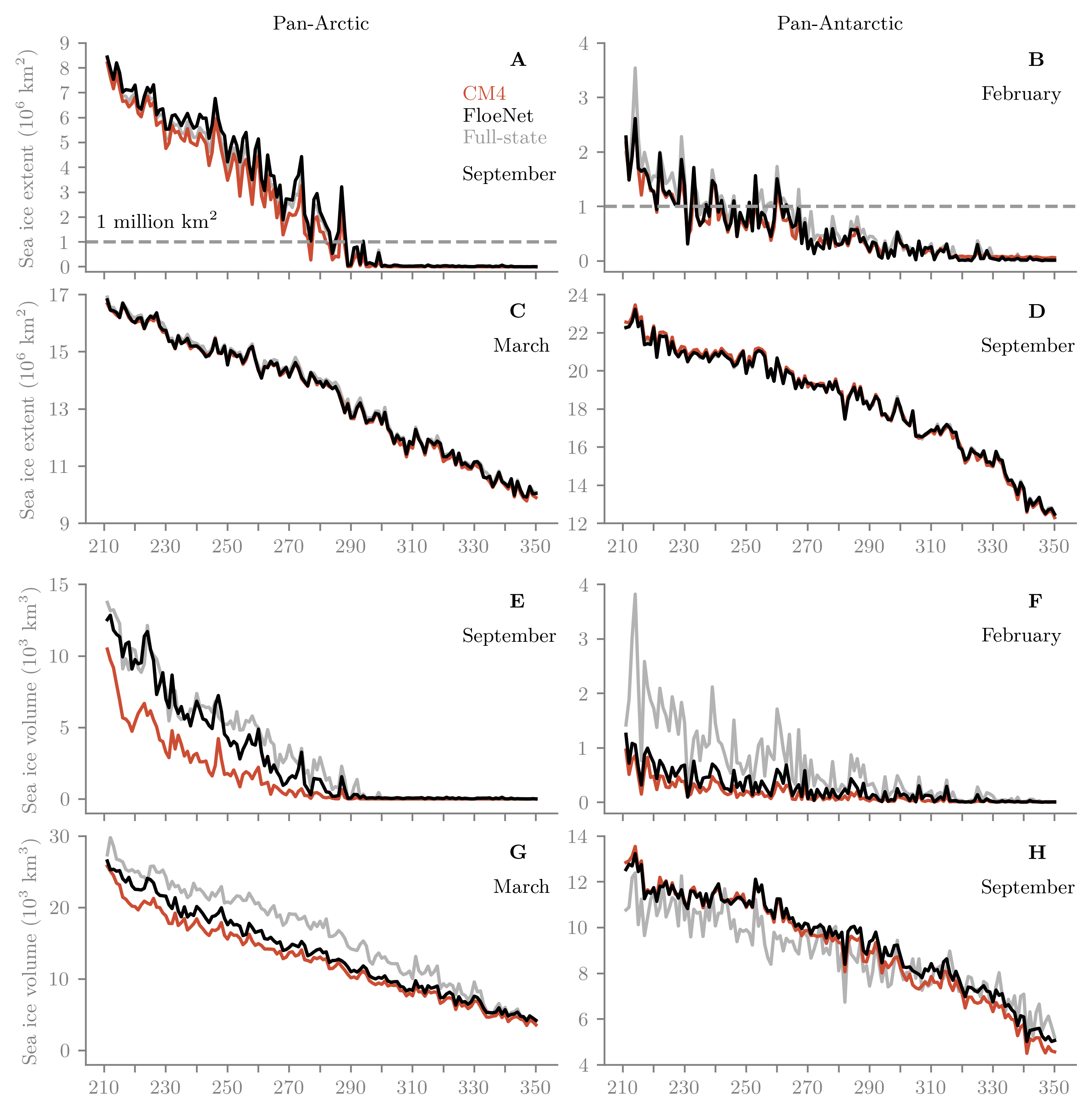}
    \caption{Summer and winter Arctic and Antarctic sea ice extent (\textbf{A}--\textbf{D}) and volume (\textbf{E}--\textbf{H}) from the 1\% CO\textsubscript{2} forcing experiment. Arctic summer and winter are given as September and March, while Antarctic summer and winter are given as February and September. These months correspond to the respective sea ice maximum and minimum extent in each hemisphere.}
    \label{fig:IceFree}
\end{figure}

\begin{figure}[t!]
    \centering
    \includegraphics[width=1\linewidth]{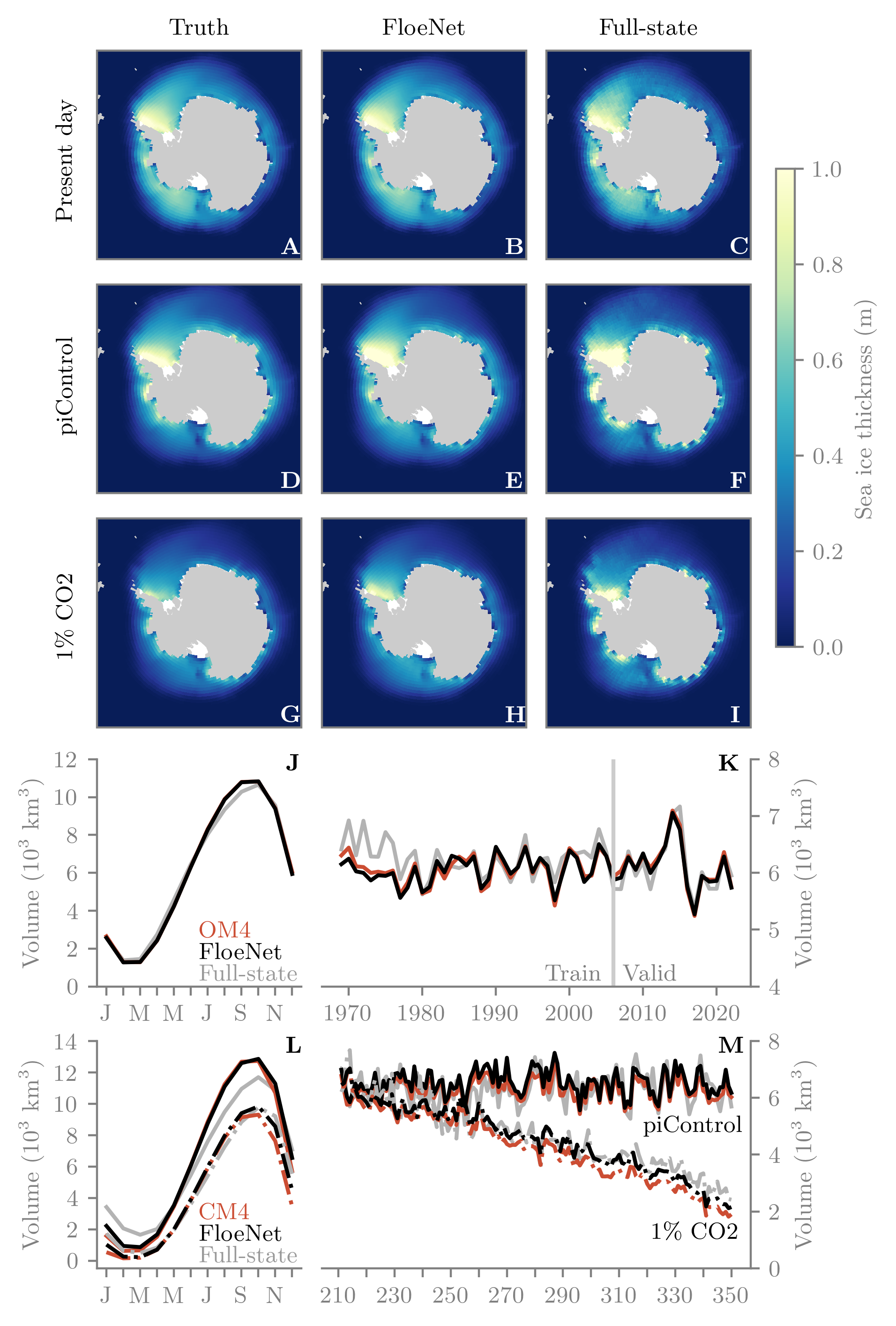}
    \caption{Same as Fig. 2 of main article, but for Antarctic sea ice thickness and volume.}
    \label{fig:AntarcticVolume}
\end{figure}

\begin{figure}[t!]
    \centering
    \includegraphics[width=1\linewidth]{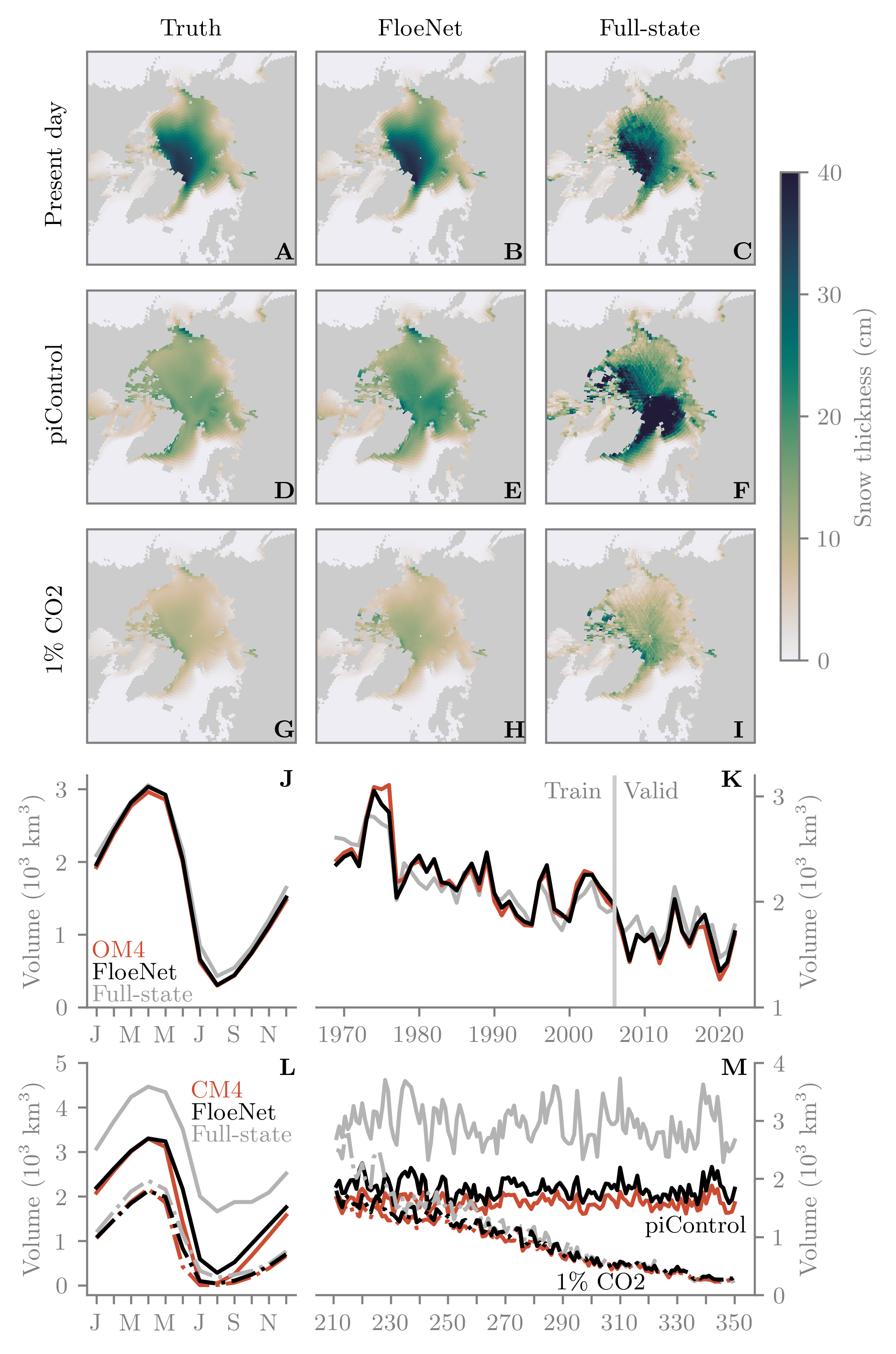}
    \caption{Same as Fig. 2 of main article, but for Arctic snow-on-sea-ice thickness and volume.}
    \label{fig:ArcticSnow}
\end{figure}

\begin{figure}[t!]
    \centering
    \includegraphics[width=1\linewidth]{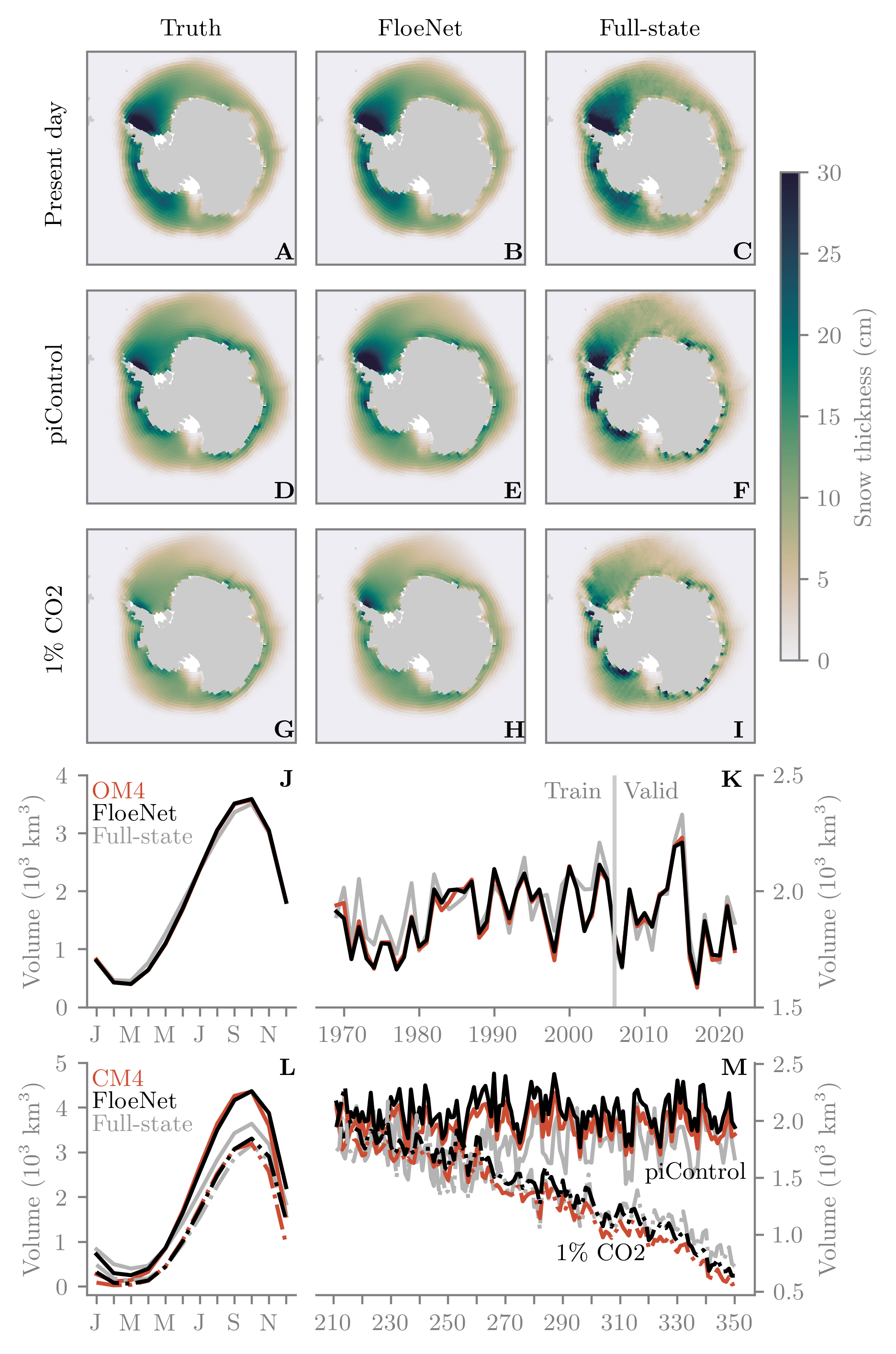}
    \caption{Same as Fig. 2 of main article, but for Antarctic snow-on-sea-ice thickness and volume.}
    \label{fig:AntarcticSnow}
\end{figure}

\begin{figure}[t!]
    \centering
    \includegraphics[width=1\linewidth]{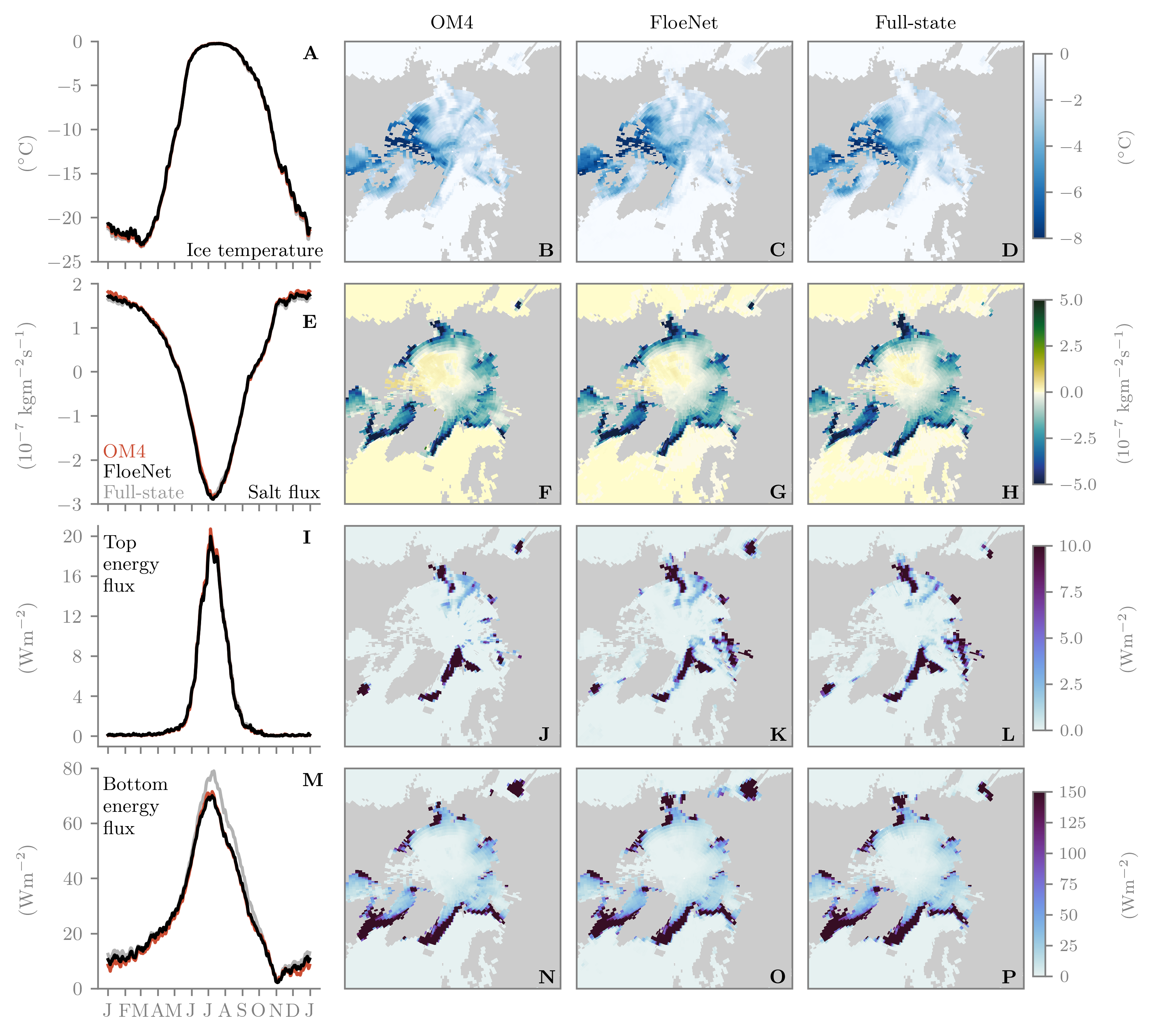}
    \caption{Evaluation of FloeNet outputs which have potential application for atmosphere-ice-ocean coupling. (\textbf{A}) Ice-surface skin temperature seasonal cycle, averaged over grid cells north of 60$^\circ$N and for the period 2006--2022. (\textbf{B}) A snapshot of ice temperature from OM4 on June 1 2018 12:00:00. (\textbf{C}) Same as (\textbf{B}) but generated from FloeNet. (\textbf{D}) Same as (\textbf{B}) but generated from the full-state model. (\textbf{E}--\textbf{H}) Same as (\textbf{A}--\textbf{D}) but for ice-to-ocean salt flux. (\textbf{I}--\textbf{L}) Same as (\textbf{A}--\textbf{D}) but for bottom melting energy flux. (\textbf{M}--\textbf{P}) Same as (\textbf{A}--\textbf{D}) but for bottom melting energy flux.}
    \label{fig:ArcticCoupling}
\end{figure}

\begin{figure}[t!]
    \centering
    \includegraphics[width=1\linewidth]{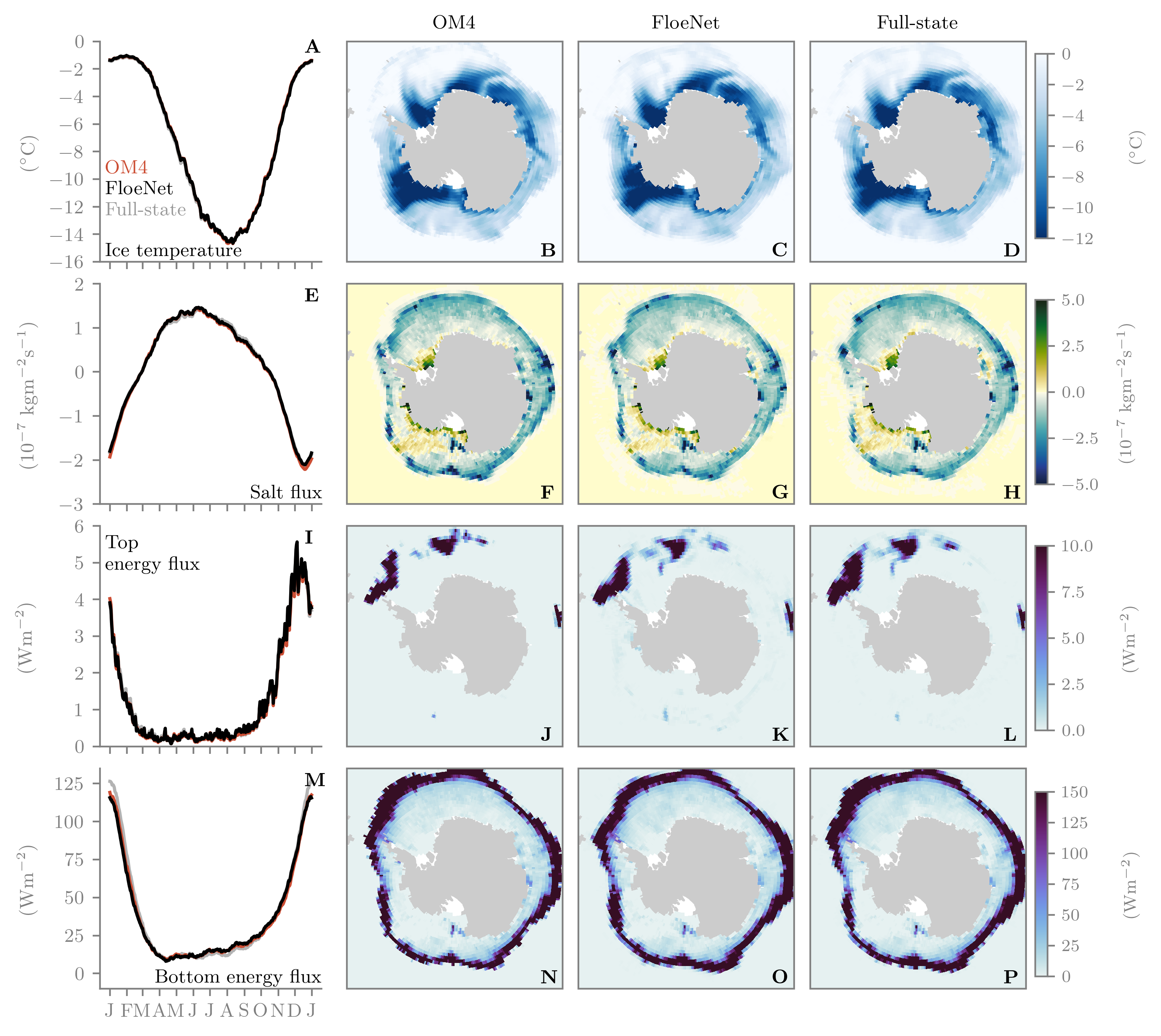}
    \caption{Same as Fig. S7, but for Antarctic. Seasonal cycles are averaged over grid cells south of 60$^\circ$S and snapshots correspond to November 1 2018 12:00:00.}
    \label{fig:AntarcticCoupling}
\end{figure}

\end{document}